\documentclass[twocolumn]{aastex631}
\usepackage{rotating}
\usepackage{graphicx}
\usepackage{supertabular}
\usepackage{longtable}
\usepackage{CJK}
\usepackage{tikz}
\usepackage{threeparttable}
\usetikzlibrary{shapes,arrows}

\received{----}
\revised{----}
\accepted{Feb 23, 2026}
\submitjournal{AJ}

\begin{document}
\begin{CJK*}{UTF8}{gbsn}
\title{The Spectroscopic and Photometric Study of a Star Cluster Sample in Andromeda Halo}

\correspondingauthor{Zhou Fan, Gang Zhao}
\email{zfan@bao.ac.cn, gzhao@nao.cas.cn}

\author[0009-0007-5610-6495]{Hongrui Gu(顾弘睿)}
\affiliation{CAS Key Laboratory of Optical Astronomy, National
  Astronomical Observatories, Chinese Academy of Sciences, Beijing
  100101, China; }
\affiliation{School of Astronomy and Space Science, University of
  Chinese Academy of Sciences, Beijing 100049, China}

\author[0000-0002-6790-2397]{Zhou Fan(范舟)}
\affiliation{CAS Key Laboratory of Optical Astronomy, National
  Astronomical Observatories, Chinese Academy of Sciences, Beijing
  100101, China; }
\affiliation{School of Astronomy and Space Science, University of
  Chinese Academy of Sciences, Beijing 100049, China}

\author[0000-0003-2472-4903]{Bingqiu Chen(陈丙秋)}
\affiliation{South-Western Institute For Astronomy Research,
  Yunnan University, Chenggong District, Kunming
  650500, China; }

\author[0000-0003-3389-2263]{Xiaoying Pang(庞晓莹)}
\affiliation{Xi'an Jiaotong-Liverpool University, 111 Ren'ai Road
Suzhou Industrial Park, Suzhou, Jiangsu Province, 215123, China; }

\author[0000-0003-3243-464X]{Juanjuan Ren(任娟娟)}
\affiliation{CAS Key Laboratory of Optical Astronomy, National
  Astronomical Observatories, Chinese Academy of Sciences, Beijing
  100101, China; }

\author[0009-0001-0604-072X]{Ruizheng Jiang(江睿铮)}
\affiliation{CAS Key Laboratory of Optical Astronomy, National
  Astronomical Observatories, Chinese Academy of Sciences, Beijing
  100101, China; }
\affiliation{School of Astronomy and Space Science, University of
  Chinese Academy of Sciences, Beijing 100049, China}

\author[0000-0003-3116-5038]{Song Wang(王松)}
\affiliation{CAS Key Laboratory of Optical Astronomy, National
  Astronomical Observatories, Chinese Academy of Sciences, Beijing
  100101, China; }
  
\author[0000-0003-0173-6397]{Kefeng Tan(谈克峰)}
\affiliation{CAS Key Laboratory of Optical Astronomy, National
  Astronomical Observatories, Chinese Academy of Sciences, Beijing
  100101, China; }

\author[0000-0002-2241-7945]{Nan Song(宋楠)}
\affiliation{China Science and Technology Museum, Beijing, 100101, China; }

\author[0000-0002-8442-901X]{Chun Li(李春)}
\affiliation{CAS Key Laboratory of Optical Astronomy, National
  Astronomical Observatories, Chinese Academy of Sciences, Beijing
  100101, China; }

\author[0000-0001-6637-6973]{Jie Zheng(郑捷)}
\affiliation{CAS Key Laboratory of Optical Astronomy, National
  Astronomical Observatories, Chinese Academy of Sciences, Beijing
  100101, China; }

\author[0000-0002-8980-945X]{Gang Zhao(赵刚)}
\affiliation{CAS Key Laboratory of Optical Astronomy, National
  Astronomical Observatories, Chinese Academy of Sciences, Beijing
  100101, China; }
\affiliation{School of Astronomy and Space Science, University of
  Chinese Academy of Sciences, Beijing 100049, China}

\author[0000-0002-9702-4441]{Wei Wang(王炜)}
\affiliation{CAS Key Laboratory of Optical Astronomy, National
  Astronomical Observatories, Chinese Academy of Sciences, Beijing
  100101, China; }
  
\author[0000-0002-8442-901X]{Yuqin Chen(陈玉琴)}
\affiliation{CAS Key Laboratory of Optical Astronomy, National
  Astronomical Observatories, Chinese Academy of Sciences, Beijing
  100101, China; }
  
\author[0000-0003-2868-8276]{Jingkun Zhao(赵景昆)}
\affiliation{CAS Key Laboratory of Optical Astronomy, National
  Astronomical Observatories, Chinese Academy of Sciences, Beijing
  100101, China; }
\affiliation{School of Astronomy and Space Science, University of
  Chinese Academy of Sciences, Beijing 100049, China}

\author[0000-0002-0389-9264]{Haining Li(李海宁)}
\affiliation{CAS Key Laboratory of Optical Astronomy, National
  Astronomical Observatories, Chinese Academy of Sciences, Beijing
  100101, China; }
\affiliation{School of Astronomy and Space Science, University of
  Chinese Academy of Sciences, Beijing 100049, China}

\author[0000-0001-7255-5003]{Yihan Song(宋轶晗)}
\affiliation{CAS Key Laboratory of Optical Astronomy, National
  Astronomical Observatories, Chinese Academy of Sciences, Beijing
  100101, China; }
\affiliation{School of Astronomy and Space Science, University of
  Chinese Academy of Sciences, Beijing 100049, China}

\author[0000-0003-2471-2363]{Haibo Yuan(苑海波)}
\affiliation{Institute for Frontiers in Astronomy and Astrophysics, Beijing Normal University, Beijing, 102206,
China; }
\affiliation{School of Physics and Astronomy, Beijing Normal University No.19, Xinjiekouwai St, Haidian
District, Beijing, 100875, China}

\author[0000-0001-7865-2648]{Ali Luo(罗阿理)}
\affiliation{CAS Key Laboratory of Optical Astronomy, National
  Astronomical Observatories, Chinese Academy of Sciences, Beijing
  100101, China; }
\affiliation{School of Astronomy and Space Science, University of
  Chinese Academy of Sciences, Beijing 100049, China}

\author[0009-0008-3430-1027]{Yujuan Liu(刘玉娟)}
\affiliation{CAS Key Laboratory of Optical Astronomy, National
  Astronomical Observatories, Chinese Academy of Sciences, Beijing
  100101, China; }
\affiliation{School of Astronomy and Space Science, University of
  Chinese Academy of Sciences, Beijing 100049, China}

\author[0000-0002-8337-4117]{Yaqian Wu(武雅倩)}
\affiliation{CAS Key Laboratory of Optical Astronomy, National
  Astronomical Observatories, Chinese Academy of Sciences, Beijing
  100101, China; }
\affiliation{School of Astronomy and Space Science, University of
  Chinese Academy of Sciences, Beijing 100049, China}



\begin{abstract}
   Halo star clusters serve as vital tracers for the formation and evolution of the Andromeda
  galaxy. In this work, we present physical parameters for 29 M31 halo star clusters, derived from a combination of spectroscopic and photometric data. Low-resolution spectra were acquired using the BFOSC spectrograph on the NAOC Xinglong 2.16-m telescope.
  For the photometric analysis, we utilized $\rm u_{SC}$ and $\rm v_{SAGE}$ bands from the SAGE survey, complemented by archival data from GALEX ($NUV$, $FUV$), PAN-STARRS ($grizy$) and the 2MASS ($JHK$). Ages and metallicities were determined via {\sc ULySS} (Vazdekis et al. and {\sc pegase-hr}) SSP model and the Bruzual \& Charlot (2003) (BC03) stellar population synthesis models. The derived parameters show good agreement with literature values. Notably, for three of these clusters, this study represents the first combined photometric and spectroscopic analysis.
\end{abstract}



\section{Introduction} \label{intro.sec}
The Local Group provides a unique laboratory for testing near-field cosmology, where halo star clusters serve as powerful probes of galactic formation and evolutionary history.
Following the Pan-Andromeda Archaeological Survey (PAndAS, \citealt{2009Natur.461...66M}), numerous giant stellar streams, such as the North West Stream (\citealt{2011ApJ...732...76R}), and other substructures have been identified in the halo of M31. Deep imaging from the Canada–France–Hawaii Telescope (CFHT), reaching limiting magnitudes of $g = 26.0$ and $i = 24.8$, has traced these structures up to distances of $\sim150$ kpc from the center of M31 (\citealt{2019MNRAS.484.1756M}). In parallel, extensive surveys have uncovered a large number of star clusters and dwarf galaxies within the M31 halo, extending to similar galactocentric radii.

Integrated light (IL) spectroscopy is recognized as a robust tool to analyze star clusters in external galaxies, enabling the derivation of key astrophysical parameters, such as age, chemical abundance (including [Fe/H]), kinematics, and mass, through comparison with stellar population synthesis models. These parameters provide critical insights into the assembly and evolutionary pathways of their host systems \citep{2019gsca.book.....S}. Alternatively, $\chi^2_{\rm min}$ fitting of spectral energy distributions (SEDs) provides an effective approach for parameter estimation using multi-band photometric data. A series of studies \citep[e.g.][]{2006MNRAS.371.1648F,2007ApJ...659..359M,2009AJ....137.4884M,2011AJ....141...86M,2012AJ....143...29M,2010AJ....139.1438W,2012AJ....144..191W}, have applied this technique to M31 star clusters using photometry from the Beijing–Arizona–Taiwan–Connecticut (BATC) system, obtained with a 60/90-cm Schmidt telescope. These works employed simple stellar population (SSP) models, particularly those of \cite{2003MNRAS.344.1000B} model (hereafter BC03) and the Galaxy Evolutionary Synthesis Models (GALEV; \citealt{2006A&A...457..467L},\citealt{2009MNRAS.396..462K}), and improved parameter constraints by incorporating broad-band $UBVRI$, 2MASS $JHK$, {\sl GALEX} NUV and FUV, and SDSS $ugriz$ photometry. Similarly, \cite{2003MNRAS.342..259D} utilized UV-to-NIR photometry from HST to derive ages, metallicities, and reddening for star clusters in NGC 3310 via SED fitting.

If halo star clusters are spatially coincident with stellar streams or other substructures in M31, their dynamical and chemical properties can shed light on past accretion events and tidal interactions between M31 and its satellite, M33. For instance, \cite{2016IAUS..312.....M} demonstrated that such substructures and the broader M31–M33 interaction history can be probed through spectroscopic follow-up of associated star clusters, using facilities such as the Xinglong 2.16-m telescope \citep{2011RAA....11.1298F,2012RAA....12..829F}, and the 6.5-m Multiple Mirror Telescope \citep{2016AJ....152..208F}. Further, \cite{2015RAA....15.1392C} and \cite{2016AJ....152...45C} analyzed M31 star clusters using low-resolution (R$\sim1800$) spectra from the Large Sky Area Multi-Object Fiber Spectroscopic Telescope (LAMOST; \citealt{2012RAA....12.1197C}; \citealt{2012RAA....12..723Z}), covering 3700 {\AA} to 9100 {\AA}, to derive radial velocities, ages, metallicities, and masses.

Previously, a variety of spectroscopic fitting techniques have been developed to improve the accuracy of stellar parameter estimation. Full-spectral fitting methods, such as those implemented in {\sc ULySS} (\citealt{2009A&A...501.1269K,2016AJ....152...45C}), utilize the entire observed spectrum, while alternative approaches rely on $\chi^2_{\rm min}$ fitting of Lick/IDS absorption-line indices (e.g., \citealt{2011RAA....11.1298F,2012RAA....12..829F,2016AJ....152...45C}). However, all such methods are inherently model-dependent, and their precision improves with the inclusion of additional observational constraints. Accordingly, combining $\chi^2_{\rm min}$ fitting to both SEDs and Lick indices has been shown to yield more robust and precise results than single-method analyses (\citealt{2016AJ....152..208F}). This hybrid strategy was successfully applied by \cite{2009gcgg.book..307L} to the globular cluster system of NGC 5128, leading to the discovery of a previously unrecognized population of intermediate-age, metal-poor clusters. While spectroscopy theoretically contains richer diagnostic information than photometry, practical limitations, particularly in flux calibration at short wavelengths (
$\lambda<4000$ {\AA}), often reduce its effective precision compared to broadband photometry. Moreover, when spectral coverage is incomplete, especially in the UV/blue regime, photometric data become essential for constraining the spectral energy distribution. As shown by \cite{2020ApJS..251...13F}, combining spectroscopy with photometry can significantly enhance the accuracy of parameter inference by anchoring the continuum shape and improving flux calibration.

In this study, we derive ages and metallicities for a sample of 29 star clusters in the M31 halo, selected from the catalog of \cite{2019MNRAS.484.1756M}. We perform joint modeling of spectroscopic and photometric data using stellar population synthesis techniques. The spectra were obtained with the BFOSC spectrograph on the Xinglong 2.16-m telescope, while the photometry combines new observations from the SAGES survey ($\rm u_{SC}$, $\rm v_{SAGES}$) with archival data from GALEX ($NUV$, $FUV$), PAN-STARRS ($grizy$), and 2MASS ($JHK$). The complete set of photometric and spectroscopic data is publicly accessible via the \href{https://nadc.china-vo.org/res/r101779/}{National Astronomical Data Center (NADC)}. To assess model dependence, we independently apply both the {\sc ULySS} and BC03 population synthesis models in the fitting procedure.
  
This paper is organized as follows. In Section \ref{sam.sec},
we describe the selection of samples. In Section \ref{sec3} we describe the
observational details for the Xinglong 2.16-m telescope, the SAGES survey and the archived data; in Section \ref{sedspecfit.sec}, we introduce the fitting
process of the Spectrum-SED fitting, based on $\chi^2_{\rm min}$
fitting with \cite{1994A&AS..106..275B} evolutionary tracks (hereafter Padova1994) and \cite{1955ApJ...121..161S} IMF of the BC03 and the {\sc ULySS} models. We briefly discuss the measurement results of the parameters and the UV-excess of the samples in Section \ref{dis.sec}. Finally, the summary
and concluding remarks are given in Section \ref{sum.sec}.

\section{The Selection of Star Cluster Sample in M31 Halo}
\label{sam.sec}
\cite{2019MNRAS.484.1756M} systematically investigated the density map of M31 GC system 
with PAndAS data. They found a correlation between the
bright substructures in the metal-poor halo field and positions of
star clusters at projected radii $R_{proj}=25 - 150$ kpc. 
Consequently, in this work, we selected relatively bright star clusters from the halo sample that are suitable for observation with the 2.16-m telescope to ensure sufficient signal-to-noise ratios (SNRs).

Table~1 lists the basic properties of the 29 M31 halo star clusters, including ID, RA, Dec, absolute magnitude (Mv), heliocentric radial velocity from \cite{2019MNRAS.484.1756M}, as well as $\rm u_{SAGES}$ and $\rm v_{SAGES}$ magnitudes from our SAGES observations. The clusters are sorted by $V$-magnitude.
\begin{figure}
    \resizebox{\hsize}{!}{\rotatebox{0}{\includegraphics{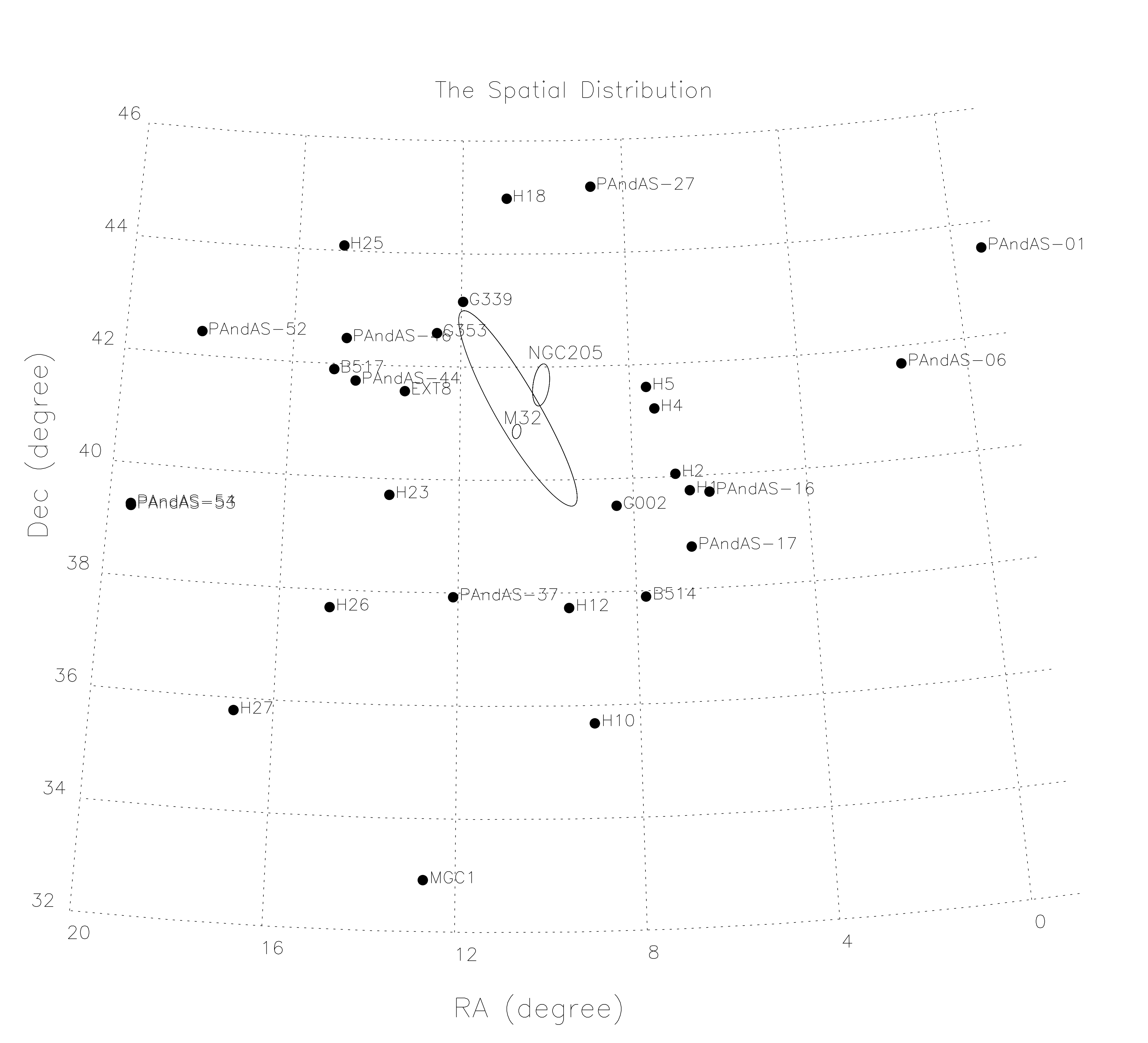}}}
    \caption{The spatial distribution of our M31 halo star cluster sample (black dots). The dashed ellipses represent the $D_{25}$ isophotes of M31, M32, and NGC 205.}
  \label{fig1}
\end{figure}
Figure~1 displays the spatial distribution of the sample star
clusters (black dots) in M31, along with M32 and NGC205 (the ellipses represent the $D_{25}$ isophotes of these galaxies). The IDs follow the nomenclature of \cite{2019MNRAS.484.1756M} and are listed in Table~1. As shown in the figure, all star clusters in our sample are located within the M31 halo.

\begin{deluxetable*}{lrrrrrr}
  \tablecolumns{7} \tablewidth{0pc} \tablecaption{Basic parameters for the sample of 29 M31 star clusters. Columns list the $\rm
    ID$, RA, Dec, absolute magnitude $M_V$ and heliocentric radial velocity from \cite{2019MNRAS.484.1756M}, 
    along with the $\rm u_{SAGES}$ and $\rm v_{SAGES}$ magnitudes from our SAGES observations.
    \label{t1.tab}}
  \tablehead{
    \colhead{$\rm ID$} & \colhead{R.A.} & \colhead{Dec} &
    \colhead{$M_V$} & \colhead{VR$_{helio}$} & \colhead{u}  & \colhead{v} \\
    \colhead{} & \colhead{(J2000)} & \colhead{(J2000)} &
    \colhead{(mag)} &\colhead{(km/s)} & \colhead{(mag)}  &
    \colhead{(mag)} } \startdata
  EXT8 & 00:53:14.5  & +41:33:24.5  &  $-9.28$  &  $-194$  &    \nodata  &    17.050  \\
MGC1 & 00:50:42.5  & +32:54:58.7  &  $-9.20$  &  $-355$  &    17.581  &    17.369  \\
PAndAS-01 & 23:57:12.0  & +43:33:08.3  &  $-7.48$  &  $-333$  &    \nodata  &    18.848  \\
B514 & 00:31:09.8  & +37:54:00.1  &  $-8.91$  &  $-471$  &    17.537  &    17.563  \\
G002 & 00:33:33.8  & +39:31:19.0  &  $-8.92$  &  $-313$  &    \nodata  &    \nodata  \\
H1 & 00:26:47.8  & +39:44:46.2  &  $-8.70$  &  $-245$  &    17.830  &    17.614  \\
H10 & 00:35:59.7  & +35:41:03.5  &  $-8.86$  &  $-358$  &    17.849  &    17.783  \\
PAndAS-46 & 00:58:56.4  & +42:27:38.3  &  $-8.67$  & $-132$  & 17.784   & 17.667    \\
PAndAS-53 & 01:17:58.4  & +39:14:53.2  &  $-9.09$  & $-271$  &    17.137  &    17.089  \\
PAndAS-54 & 01:18:00.1  & +39:16:59.9  &  $-8.58*$  &  $-345$  &    17.708  &    17.558  \\
B517 & 00:59:59.9  & +41:54:06.8  &  $-8.17$  &  $-277$  &    18.337  &    17.956  \\
H12 & 00:38:03.9  & +37:44:00.2  &  $-8.19$  &  $-396$  &    18.218  &    18.097  \\
H23 & 00:54:25.0  & +39:42:55.7  &  $-8.09$  &  $-373$  &    18.345  &    18.882  \\
H27 & 01:07:26.3  & +35:46:48.4  &  $-8.39$  &  $-291$  &    18.341  &    18.044  \\
H5 & 00:30:27.3  & +41:36:19.5  &  $-8.44$  &  $-392$  &    \nodata  &    17.754  \\
PAndAS-16 & 00:24:59.9  & +39:42:13.1  &  $-8.44$  &  $-490$  &    \nodata  &    \nodata  \\
PAndAS-17 &  00  26  52.2  &  +38  44  58.1  & $-8.17$ & $-260$ & \nodata  &    \nodata \\
H18 & 00:43:36.1  & +44:58:59.3  &  $-8.09$  & $-206$  &    18.499  &    18.428  \\
H25 & 00:59:34.6  & +44:05:38.9  &  $-7.93$  &  $-204$  &    18.820  &    18.606  \\
H4 & 00:29:45.0  & +41:13:09.4  &  $-7.82$  &  $-368$  &    \nodata  &    18.906  \\
PAndAS-06 & 00:06:12.0  & +41:41:21.0  &  $-8.02$  &  $-341$  &    18.386  &    18.414  \\
PAndAS-27 & 00:35:13.5  & +45:10:37.9  &  $-7.69$ & $-46$  & \nodata & 19.041 \\
PAndAS-44 & 00:57:55.9  & +41:42:57.0  & $-7.72$  & $-349$ & 18.714 & 18.500 \\
G339 & 00:47:50.2  & +43:09:16.5  &  $-7.58$  &  $-97$  &    19.276  &    19.016  \\
G353 & 00:50:18.2  & +42:35:44.2  &  $-7.60$  &  $-295$  &    19.061  &    18.855  \\
H2 & 00:28:03.2  & +40:02:55.6  &  $-7.50$  &  $-519$  &    18.976  &    18.822  \\
PAndAS-37 & 00:48:26.5  & +37:55:42.1  &  $-7.35$ & $-404$ & 19.345 & 19.100 \\ 
PAndAS-52 & 01:12:47.0  & +42:25:24.9  &  $-7.58$  &  $-297$  &    18.257  &    19.049  \\
H26 & 00:59:27.5  & +37:41:30.9  &  $-7.40$  &  $-411$  &    19.569  &    19.293  \\
  \enddata
\end{deluxetable*}
\section{Observations and Data Reduction}\label{sec3}

\subsection{The Spectroscopic Observations}
Low-resolution spectroscopic observations were carried out in
September and November 2019 using the Beijing Faint Object Spectrograph and Camera (BFOSC) on the Xinglong 2.16-m reflector
\citep{2016PASP..128k5005F}. The telescope is located at the Xinglong Observatory of the National Astronomical Observatories, Chinese Academy of Sciences (NAOC). All 29 clusters were exposed for 3600 seconds 
(Table~2). Weather conditions were clear for most of the observing nights, with typical seeing of $\sim2''$, a temperature of 20$^\circ C$, and relative humidity of 30\%. We used Grism G4 with a slit width of $1.8''$. This configuration provides a wavelength coverage of 3850-7000 {\AA} with a first-order dispersion of 4.45 {\AA} pixels$^{-1}$. The instrument has a nominal spectral resolution of $R=620$ for a $0.6''$ slit at a central wavelength of 5007 {\AA} \citep{2016PASP..128k5005F}. 
The observational information of our sample is listed
in Table~2, including cluster IDs (following \citealt{2019MNRAS.484.1756M}), observation dates, times, exposure times and slit configurations. 

Table~3 lists the parameters for the Andor BEX2-DD Camera mounted on BFOSC, including the A/D rate, readout time (ROT), pre-amp, gain, and readout noise (RON) for the High-Sensitivity and High-Capacity modes. The instrument is equipped with an Andor BEX2-DD iKon-L 936 camera, featuring a large-area (2048$\times$2048 pixel) back-illuminated E2V CCD42-40 NIMO sensor.
This high-dynamic-range CCD utilizes the 'Dual AR Extended Range' sensor option, providing broad spectral sensitivity from the ultraviolet (UV) to the near-infrared (NIR).
The quantum efficiency (QE) exceeds 90\% across the wavelength range of approximately 4000–8500 {\AA} (see Figure~2). Each pixel has a physical size of 13.5 $\mu$m, and the sensor is thermo-electrically (TE) cooled to $-100 ^\circ C$. In High Sensitivity mode with 1MHz readout at 16-bit depth and 2$\times$ binning, the system achieves a gain of 1.41 e$^{-}$ ADU$^{-1}$ and a readout noise of 4.64 e$^{-}$.

\subsection{Spectral data Reduction}
\begin{figure}
  \centering
      \includegraphics[angle=0,scale=0.06222]{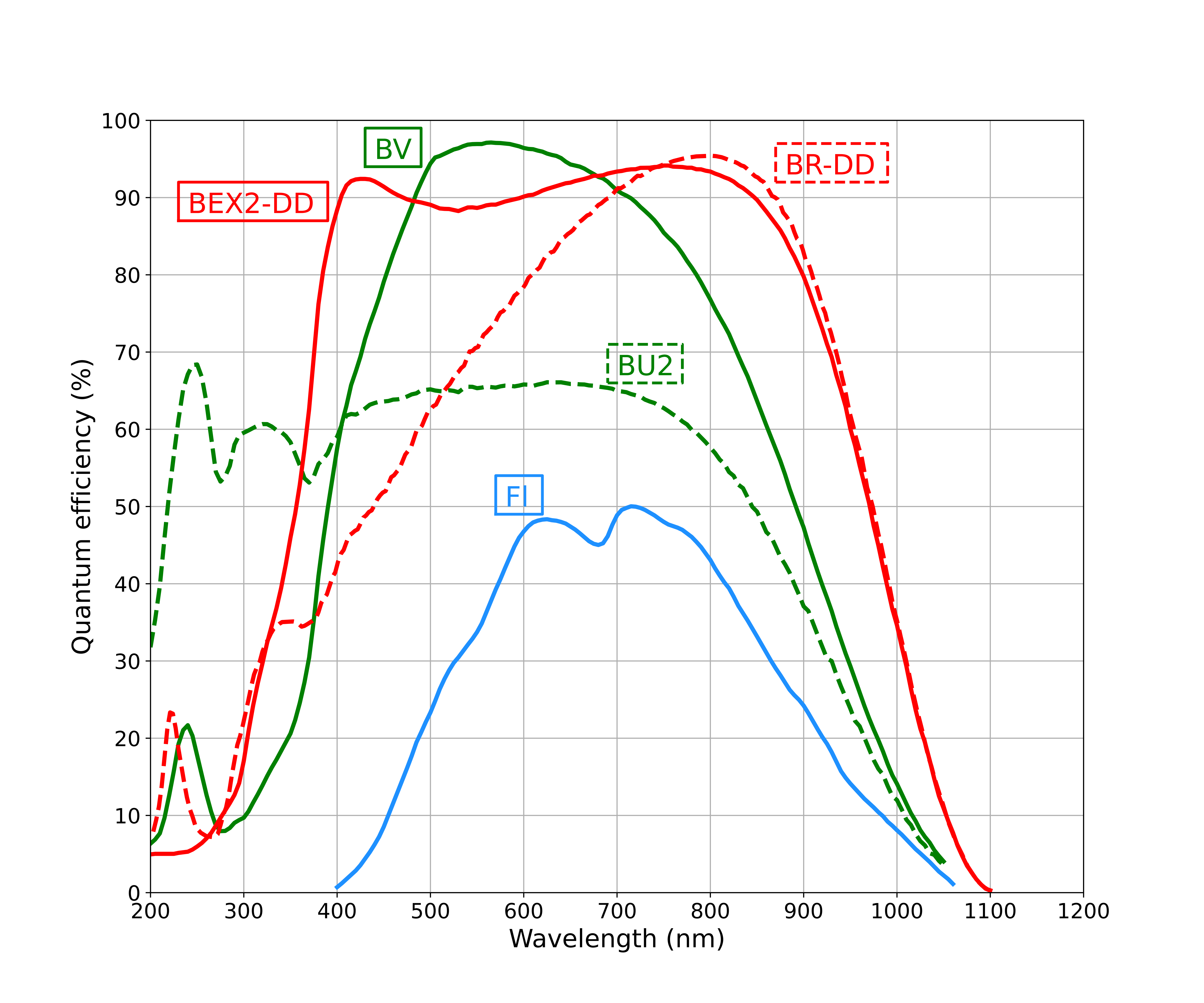}	
    \caption{Quantum efficiency (QE) curve of the Andor iKon-L 936 BEX2-DD
      CCD mounted on the BFOSC instrument (red solid line). The QE exceeds 90\% across most of
      the optical wavelength range.} 
  \label{fig1a}
\end{figure}
The data reduction followed the standard procedures with the NOAO
Image Reduction and Analysis Facility ({\sc iraf} v.2.15) software
package. After a visual inspection of the spectral image, bias frames were combined using {\tt zerocombine} and applied via {\tt ccdproc}. Flat-field frames were combined,
normalized, and applied using {\tt flatcombine}, {\tt
response}, and {\tt ccdproc}. Cosmic rays were removed using the {\tt cosmicrays} package. The star cluster and comparison arc lamp spectra were extracted using {\tt apall}. Wavelength calibration was performed using helium/argon lamp spectra taken at the beginning and end of each observing night. Spectral features in the comparison lamps were identified with {\tt
identify}, and the wavelength solution was applied using {\tt refspectra}. The spectra were then dispersion-corrected and resampled using {\tt dispcor}. For flux calibration, we used four Kitt Peak
National Observatory (KPNO) spectral standard stars from \cite{1988ApJ...328..315M}. The {\tt standard} and {\tt
  sensfunc} packages were used to combine the standard stars and
determine the sensitivity and extinction of the atmosphere. Finally, the {\tt calibrate} package was applied to correct for extinction and complete the flux calibration.

\begin{deluxetable*}{lcccc}
  \tablecolumns{5} \tablewidth{0pc} \tablecaption{Observation
    information for our sample of M31 star clusters. Columns list the ID, RA, Dec
    and $V$-band magnitude.   
    \label{t3.tab}}
  \tablehead{
    \colhead{$\rm ID$} & \colhead{Date} & \colhead{Time} &
    \colhead{Expose Time} & \colhead{Grism/Slit width}  \\
    \colhead{} & \colhead{(yyyymmdd)} & \colhead{(Beijing)} & \colhead{(second)}
    &\colhead{} } \startdata
    EXT8 & 20190921 & 22:01:06 & 3600 & G7+S1.8+385LP \\
MGC1 & 20190921 & 23:14:12 & 3600 & G7+S1.8+385LP \\
PAndAS-01 & 20190921 & 20:55:37 & 3600 & G7+S1.8+385LP \\
B514 & 20190922 & 22:09:38 & 3600 & G7+S1.8+385LP \\
G002 & 20190922 & 21:01:11 & 3600 & G7+S1.8+385LP \\
H1 & 20190922 & 25:41:20 & 3600 & G7+S1.8+385LP \\
H10 & 20190922 & 24:20:48 & 3600 & G7+S1.8+385LP \\
PAndAS-46 & 20190922 & 26:47:49 & 3600 & G7+S1.8+385LP \\
PAndAS-53 & 20190922 & 23:15:23 & 3600 & G7+S1.8+385LP \\
PAndAS-54 & 20190922 & 27:52:27 & 3600 & G7+S1.8+385LP \\
B517 & 20190923 & 25:41:03 & 3600 & G7+S1.8+385LP \\
H12 & 20190923 & 24:20:26 & 3600 & G7+S1.8+385LP \\
H23 & 20190923 & 27:50:25 & 3600 & G7+S1.8+385LP \\
H27 & 20190923 & 23:14:15 & 3600 & G7+S1.8+385LP \\
H5  & 20190923 & 20:59:36 & 3600 & G7+S1.8+385LP \\
PAndAS-16 & 20190923 & 22:07:50 & 3600 & G7+S1.8+385LP \\
PAndAS-17 & 20190923 & 26:45:15 & 3600 & G7+S1.8+385LP \\
H18 & 20190924 & 21:14:45 & 3600 & G7+S1.8+385LP \\
H25 & 20190924 & 23:33:34 & 3600 & G7+S1.8+385LP \\
H4 & 20190924 & 24:43:35 & 3600 & G7+S1.8+385LP \\
PAndAS-06 & 20190924 & 22:23:05 & 3600 & G7+S1.8+385LP \\
PAndAS-27 & 20190924 & 27:35:46 & 3600 & G7+S1.8+385LP \\
PAndAS-44 & 20190924 & 26:13:47 & 3600 & G7+S1.8+385LP \\
G339 & 20190925 & 25:14:28 & 3600 & G7+S1.8+385LP \\
G353 & 20190925 & 22:10:07 & 3600 & G7+S1.8+385LP \\
H2 & 20190925 & 26:38:24 & 3600 & G7+S1.8+385LP \\
PAndAS-37 & 20190925 & 27:46:19 & 3600 & G7+S1.8+385LP \\
PAndAS-52 & 20190925 & 23:19:38 & 3600 & G7+S1.8+385LP \\
H26 & 20191111 & 18:55:51 & 3600 & G7+S1.8+385LP \\
    \enddata
  \end{deluxetable*}
\begin{deluxetable*}{ccc|cc|cc}
  \turnpage
  \tablecolumns{7} \tablewidth{0pc} \tablecaption{Gain, readout
    noise (RON), and readout time (ROT) for the BFOSC camera on the Xinglong 2.16-m telescope. Estimates are based on the central 200 x 200 pixels region ([901:1100,901:1100]). Parameters are listed for both the High-Sensitivity and High-Capability modes. 
    \label{t2.tab}}
  \tablehead{\colhead{A/D rate} & \colhead{ROT} &
    \colhead{Pre-amp} & \multicolumn{2}{c}{Gain} &\multicolumn{2}{c}{RON} \\
    \cline{4-7}
    \colhead{(Mbit)} & \colhead{(seconds)} &   \colhead{} &
    \colhead{High-Sen} &  \colhead{High-Cap} & \colhead{High-Sen} &  \colhead{High-Cap}}
  \startdata
  5&	   1&	$\times1$&	4.03&	14.6&	23.68&	76.8\\
5&	   1&	$\times2$&	2.18&	8.15&	18.11&	60.09\\
5&	   1&	$\times4$&	1.14&	4.28&	14.19&	42.53\\
3&	   2&	$\times1$&	2.92&	10.92&	14.73&	54.19\\
3&     2&	$\times2$&	1.53&	5.92&	8.88&	33.95\\
3&	   2&	$\times4$&	0.80&	2.95&	7.69&	23.79\\
1&	   5&	$\times1$&	2.64&	10.44&	6.09&	24.67\\
1&	   5&	$\times2$&	1.41&	5.81&	4.64&	17.22\\
1&	   5&	$\times4$&	0.78&	3.07&	4.60&	15.08\\
0.05&	90&	$\times1$&	2.63&	10.59&	2.98&	9.86\\
0.05&	90&	$\times2$&	1.40&	5.61&	2.48&	7.27\\
0.05&	90&	$\times4$&	0.76&	2.97&	2.34&	6.29\\
  \enddata
\end{deluxetable*}

\subsection{Photometric data Reduction}
\subsubsection{Archived data}
We downloaded science images from the \href{http://ps1images.stsci.edu/cgi-bin/ps1cutouts?pos=&filter=color&filetypes=stack&auxiliary=data&size=6000&output_size=1024&verbose=0&autoscale=99.500000&catlist=}{PAN-STARRS}, \href{https://irsa.ipac.caltech.edu/applications/2MASS/IM/interactive.html}{2MASS}, and \href{https://mast.stsci.edu/portal/Mashup/Clients/Mast/Portal.html}{GALEX} surveys based on the coordinates in Table~1 and performed aperture photometry for each star cluster. We measured the fluxes within different apertures and constructed growth curves of each cluster. The aperture yielding the highest signal-to-noise ratio was selected as the optimal aperture, 
while the asymptotic value of the growth curve was adopted as the total flux. 
By calculating the ratio between the optimal aperture flux and the total flux, we applied aperture corrections to obtain the final flux and associated uncertainty for each cluster in each band.

For PAN-STARRS and 2MASS images, the header files provide the magnitude zero-points, allowing direct conversion of measured fluxes to magnitudes. For GALEX images, although the zero-points are not included in the headers, the corresponding source catalogs provide both flux and magnitude values for detected objects. We used these to derive the zero-points and applied them to convert our measured fluxes into magnitudes.

The calculation of magnitude uncertainties was based on the background noise from the sky annulus and photon shot noise within the aperture. However, this theoretical noise estimation method tends to underestimate the true noise level. Consequently, we compared our computed uncertainties with those reported in the official catalogs and derived a correction factor for each band. These factors were then applied to rescale our uncertainties, ensuring they more accurately reflect the true photometric scatter.

\begin{figure*}
  \centering
      \includegraphics[angle=0,scale=0.1]{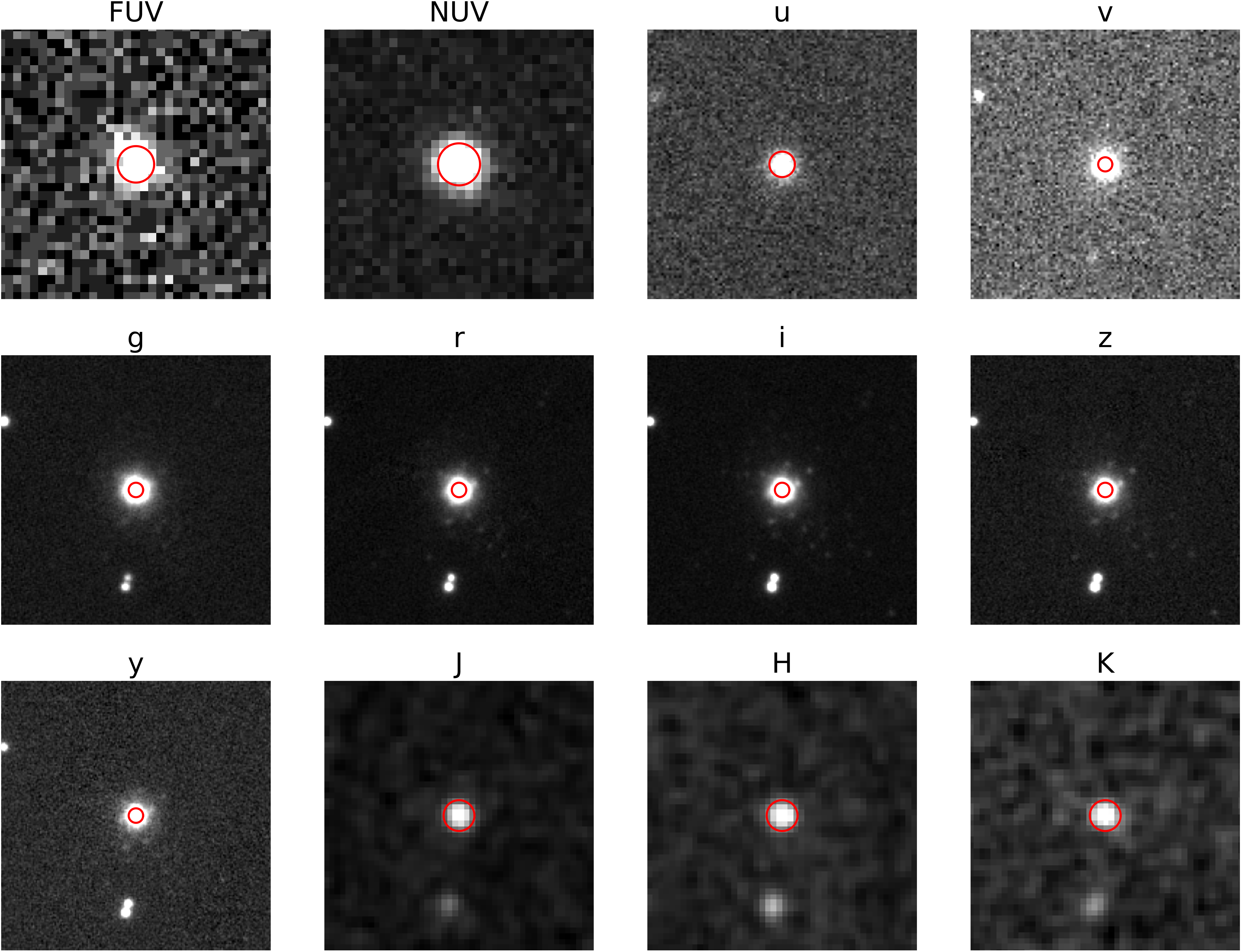}	
    \caption{Cutout images of star cluster G002 in 12 filters: GALEX(NUV, FUV); SAGES ($\rm u_{SC}$,$\rm v_{SAGES}$); Panstarrs ($g,r,i,z,y$), and 2MASS ($J,H,K$). Each panel has a field of view of 48" $\times $ 48". The red circle indicate the apertures used for photometry, optimized for signal-to-noise ratio.
    } 
  \label{fig1a}
\end{figure*}

Figure~3 displays the images of star cluster G002 across different bands, with a field of view of 48″ $\times$ 48″. The red circles indicate the apertures with optimal signal-to-noise ratio for photometry in each band.
\subsubsection{SAGES data}
The SAGES
survey\footnote{http://sage.sagenaoc.science/~sagesurvey/} \citep[PI: Gang Zhao, see e.g.][]{2018RAA....18..147Z,2019RAA....19....3Z,2023ApJS..268....9F} began operations in 2015, covering 12,000 deg$^2$ of the northern sky with
declination $\delta
>-5^{\circ}$ and excluding the bright, high-extinction Galactic disk ($|b|<10^{\circ}$). The survey provides photometry in
eight bands $\rm u_{SC}$,$\rm v_{SAGES}$, $g$, $r$, $i$, $\rm
H{\alpha}_n$, $\rm H{\alpha}_w$ and DDO51. The central wavelengths
of the $\rm u_{SC}$ and $\rm v_{SAGES}$ bands are 3520 and 3950 {\AA}
respectively. Blue-band filters often encompass various metallicity-sensitive absorption lines. In 
particular, the $\rm v_{SAGES}$ band covers the Ca\,II K line at
$\lambda=3933.44$ {\AA} (between H$\epsilon$ and H$\zeta$), which is
highly sensitive to metallicity for FGK stars. As observations of the
$\rm u_{SC}$ and $\rm v_{SAGES}$ bands are nearly complete ($\sim88$\% ), we utilize photometry from these two bands to constrain the ages and metallicities of our M31 star cluster sample. Observations for these two bands were conducted using the 90-inch (2.3-m) Bok
telescope of Steward Observatory, University of Arizona. A CCD
mosaic camera, which consists of four 4k$\times$4k CCDs, is mounted
at the prime focus. The field of view is $1\time1$ deg$^2$ and the
pixel size is $0.45''$. The photometry pipeline is based on the {\sc
SExtractor} and MAG$\_$AUTO. For astrometry, the {\sc SCAMP} was used and the Position and Proper Motion Extended (PPMX, \citealt{2008A&A...488..401R}) catalog was adopted in our pipeline as the astrometric
reference. The detailed description of the pipeline of photometry,
astrometry, and flux calibrations can be found in \cite{2019RAA....19....3Z}. A SNR of 100 corresponds to a limiting magnitude of
$\rm u_{SC}\sim16.5$ mag and $\rm v_{SAGES}\sim15.5$ mag; 5 to $\rm
u_{SC}\sim20$ mag and $\rm v_{SAGES}\sim19.5$ mag. 
For our analysis, since star clusters are extended sources, we constructed growth curves to calculate aperture corrections for the clusters in both the 
$\rm u_{SC}$, $\rm v_{SAGES}$ bands.

\section{Fitting with spectroscopy and the photometry of GALEX, SAGES, UBVRI, 
ugriz and 2MASS}
\label{sedspecfit.sec}
In ground-based observations, M31 star clusters appear as point-like or slightly extended sources. Since the atmospheric seeing generally exceeds the clusters' angular diameters. Consequently, whether for spectral analysis or photometric measurement, and regardless of the photometric aperture size or spectral slit width, the observed light originates from the entire star cluster, effectively sampling the same stellar population. For this reason, we can jointly fit spectroscopic and photometric data in our subsequent analysis. 

Initially, we applied the BC03 models to fit the photometric and spectroscopic data. However, due to the low spectral resolution of the BC03 templates, precise abundance measurements are difficult. To better assess the BC03 results and obtain more accurate abundance estimates, we also employed the {\sc ULySS} \citep{2009A&A...501.1269K} model to perform full spectral fitting of the spectroscopic data, from which we derived the age and metallicity of the star clusters.

\subsection{Fitting with BC03}

The evolutionary stellar population synthesis
models of BC03 not only provide spectra
and SEDs for different physical parameters, but also Lick/IDS
absorption-line indices. The models adopt Padova1994 stellar evolutionary tracks, with initial mass functions (IMFs)
of \cite{1955ApJ...121..161S}. The wavelength coverage ranges
from 91 {\AA} to 160 $\mu$m. The Padova1994 offers six
metallicity options ($Z=0.0001$, 0.0004, 0.004, 0.008, 0.02, and
0.05) and there are 221 age steps
from 0 to 20 Gyr in total. As the resolution of metallicity grid is insufficient for precise fitting, we interpolated the model grid from the original six metallicity steps to 61 values to ensure smoother fitting.

In this section, we gathered the photometry of our sample star clusters in the 
$GALEX$ NUV, FUV, $\rm u_{SC}$ (SAGES), $\rm v_{SAGES}$ (SAGES), PAN-STARRS $grizy$ and the 2MASS $JHK$ bands. All photometric magnitude measurements are described in Section 3.  
Since different photometric magnitudes were given in different magnitude systems (Vega and AB), we adopted the data from \citet{2018ApJS..236...47W} and applied corrections of 0.87, 1.344, and 1.814 to the J, H, K-bands, respectively, to convert the 2MASS Vega magnitudes to the AB system, ensuring consistency with the other photometric systems used in this work.

Extinction correction is also a crucial step prior to spectral fitting. We adopted the E(B-V) values provided by M31 Revised Bologna Clusters and Candidates Catalog Version 5 (\citealt{2014yCat.5143....0G}) as the extinction values for the clusters; these values are the averages of those given in \cite{2000AJ....119..727B} and \cite{2011AJ....141...61C}. For clusters not included in this catalog, we used the extinction estimates from the \cite{1998ApJ...500..525S} (here after SFD98) dust maps. While SFD98 values are generally less precise than spectroscopic derivations, they are acceptable for our analysis because the M31 halo has low extinction and the dust emission is not saturated, ensuring a reliable relationship between emission and extinction. We also applied the recalibration factor of 0.86 recommended by \citealt{2011ApJ...737..103S}. Finally, we corrected both the spectroscopic and photometric data for reddening using the extinction law of \cite{1989ApJ...345..245C} (Equations 6-7).

Following these corrections, the photometric and spectroscopic data were fitted simultaneously. The photometric and spectroscopic data were fitted simultaneously using $\chi^2$ minimization.

\begin{equation}
\chi^2_{\rm min}={\rm min}\left(\chi^2_{\rm spec}+\sum_{j=1}^{n_{phot}}\chi^2_{\rm phot_j}\right),
\end{equation}

$n_{phot}$ is the number of photometric bands.

\begin{equation}
\chi^2_{\rm spec}=\sum_{i=1}^{n_{spec}}\left({\frac{m_{\lambda_i}^{\rm
            obs}-m_{\lambda_i}^{\rm mod}(t,\rm [Z/H])}
        {\sigma_{m,i}}}\right)^2,
\end{equation}

In equation~2, $n_{spec}$ is the number of spectral data points (pixels); $m_{\lambda_i}^{\rm obs}$ is the AB magnitude that is
transformed from the dereddened observed spectra;
$m_{\lambda_i}^{\rm mod}(t,\rm [Z/H])$ is the $i^{\rm th}$ magnitude provided in the
stellar population model at an age $t$ and metallicity $\rm [Z/H]$. $\sigma_{m,i}$ represent the observation uncertainties in the spectroscopy. 

\begin{equation}
\chi^2_{\rm phot_j}=\left({\frac{M_{\lambda_j}^{\rm
            obs}-M_{\lambda_j}^{\rm mod}(t,\rm [Z/H])}
        {\sigma_{M,j}}}\right)^2\times k_j,
\end{equation}

Similarly, in equation~3, $M_{\lambda_j}^{\rm obs}$ represents the observed
dereddened magnitude in the $j^{\rm th}$ band, including $\rm u_{SC}$ (SAGES), $\rm v_{SAGES}$ (SAGES), $NUV$ (GALEX), $FUV$ (GALEX), $grizy$ (PAN-STARRS), $JHK$ (2MASS) bands; $M_{\lambda_j}^{\rm
mod} (t,\rm [Z/H])$ is the fitted $j^{\rm th}$ magnitude from the
stellar population model at an age $t$, metallicity $\rm [Z/H]$; and $\sigma_{M,i}$ represent the observation uncertainties in the photometry. $k_j$ is the weight of the $j^{\rm th}$ filter while fitting.

\begin{equation}
k_j=\sqrt{\frac{R_{spec}\cdot W_{filter}}{\lambda _{filter}}}.
\end{equation}

While a single photometric filter provides only one data point, 
its bandpass is much broader than a single spectral resolution element, meaning it aggregates more signal.
However, spectroscopy intrinsically contains higher information density per unit wavelength coverage. To balance these differences, we defined the weights $k_j$ as shown in Equation (4).

\begin{equation}
\chi ^2 _{final}= \frac{\chi ^2 _{min}}{n_{spec}+\sum_{j=1}^{n_{phot}}k_j}
\end{equation}

\begin{equation}
{\rm exp} \left( \frac{1}{\chi ^2 _{min} + \Delta \chi ^2}\right)=0.5 \cdot {\rm exp} \left( \frac{1}{\chi ^2 _{min}}\right)
\end{equation}

Since $\chi ^2$ scales with the number of data points, it cannot directly compare fits with different sample sizes. We therefore used the reduced chi-square (Equation 5) to assess fit quality. Parameter uncertainties were determined using the $\Delta \chi ^2$ criterion (Equation 6), where the $1 \sigma$ (68\%) confidence interval is defined by $\chi ^2 \leq \chi ^2 _{min} + \Delta \chi ^2$. 
Figure~4 illustrates the joint fitting result for cluster G002, with the upper panel showing the best-fit model and data, and the lower panel displaying the residuals. The derived ages and metallicities for all 29 star clusters using the BC03 are listed in Table~4.

\begin{figure*}
  \centering
      \includegraphics[angle=0,scale=0.1111]{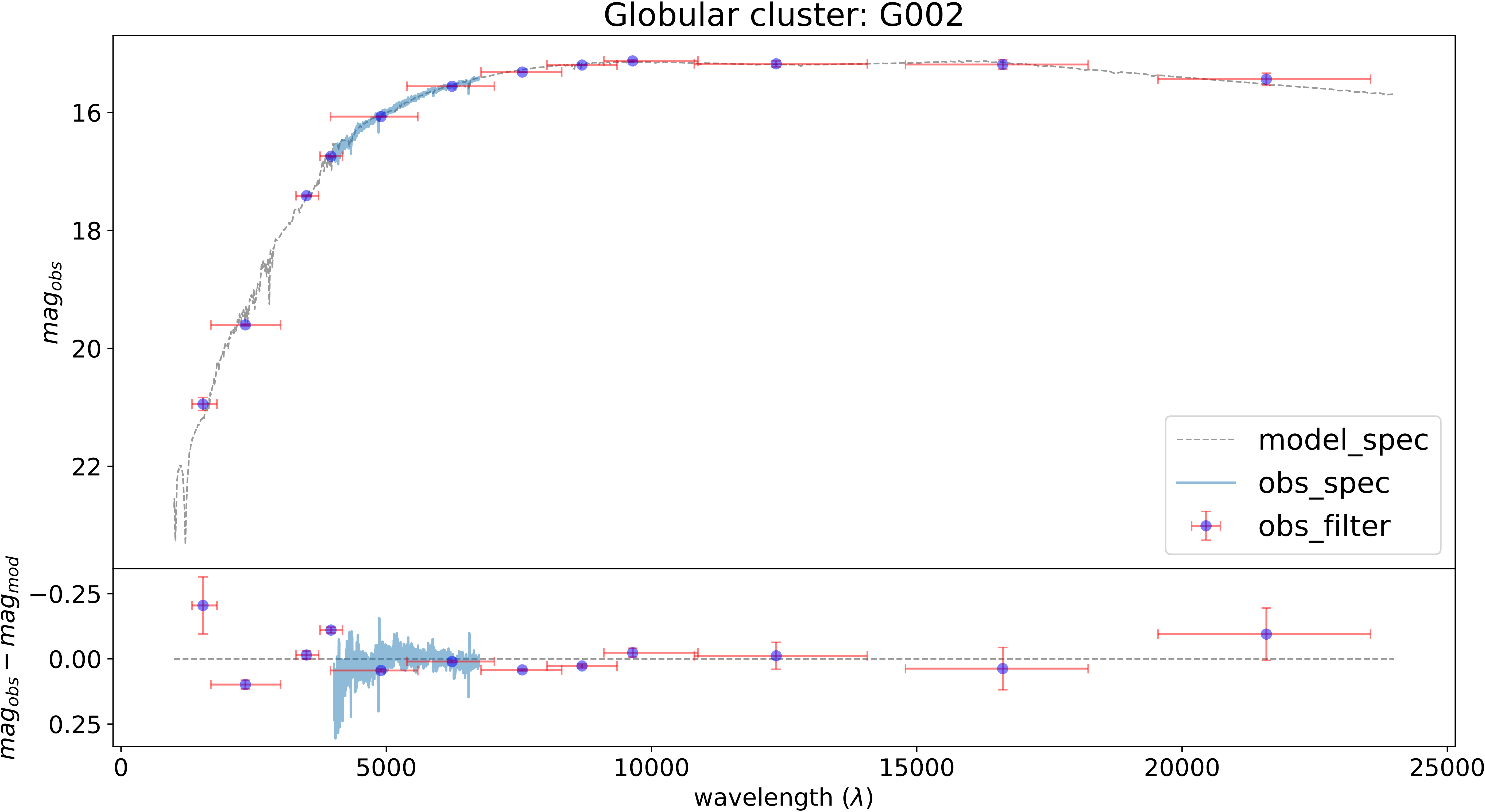}	
    \caption{Joint spectral and photometric fitting for star cluster G002 using BC03 models.\textbf{Top panel:} The observed spectrum (blue solid line) and photometry (blue points) overlaid with the best-fit model (gray dashed line). Horizontal error bars on photometric points represent the filter bandwidths, while vertical bars indicate photometric uncertainties.\textbf{Bottom panel:} The fitting residuals. 
    } 
  \label{fig1a}
\end{figure*}

\begin{deluxetable}{r|rrr}
  \tablecolumns{4} \tablewidth{0pc} \tablecaption{Fitting results derived from the joint analysis of spectroscopy and all available photometry. We used the BC03 models based on Padova1994 evolutionary tracks and a Salpeter IMF \citep{1955ApJ...121..161S}
  }
    \label{t5.tab}
  \tablehead{\colhead{}  &\multicolumn{3}{c}{\cite{1955ApJ...121..161S}+P1994} \\
    \cline{2-4}
    \colhead{ID} & \colhead{log $t$} &  \colhead{$\rm
      [Fe/H]$} &  \colhead{$\chi^2_{min}/dof$}  \\
    \colhead{} & \colhead{(yr)} & \colhead{(dex)} & \colhead{} } 
  \startdata
EXT8 & $9.676_{ -0.144}^{+0.065}  $ & $   -1.506_{-0.213}^{+0.211} $ & 4.9436\\
MGC1 & $10.196_{ -0.13}^{+0.096}  $ & $   -1.359_{-0.25}^{+0.187} $ & 2.7765\\
PAndAS-01 & $9.988_{ -0.432}^{+0.312}  $ & $   -1.506_{-0.734}^{+0.766} $ & 0.9719\\
B514 & $10.3_{ -0.052}^{+0.0}  $ & $   -1.506_{-0.3}^{+0.167} $ & 13.7689\\
G002 & $10.196_{ -0.053}^{+0.061}  $ & $   -1.946_{-0.093}^{+0.086} $ & 26.8766\\
H1 & $9.467_{ -0.067}^{+0.167}  $ & $   -0.772_{-0.231}^{+0.157} $ & 4.892\\
H10 & $9.78_{ -0.191}^{+0.078}  $ & $   -0.918_{-0.237}^{+0.129} $ & 2.8562\\
PAndAS-46 & $10.3_{ -0.057}^{+0.0}  $ & $   -1.799_{-0.272}^{+0.279} $ & 7.6289\\
PAndAS-53 & $10.3_{ -0.053}^{+0.0}  $ & $   -2.093_{-0.147}^{+0.216} $ & 18.979\\
PAndAS-54 & $9.571_{ -0.14}^{+0.087}  $ & $   -0.625_{-0.14}^{+0.165} $ & 30.2655\\
B517 & $9.78_{ -0.315}^{+0.127}  $ & $   -0.918_{-0.299}^{+0.223} $ & 6.8321\\
H12 & $9.78_{ -0.228}^{+0.176}  $ & $   -1.506_{-0.596}^{+0.337} $ & 1.2732\\
H23 & $10.092_{ -0.195}^{+0.208}  $ & $   -0.918_{-0.377}^{+0.252} $ & 2.8819\\
H27 & $10.3_{ -0.054}^{+0.0}  $ & $   -1.653_{-0.306}^{+0.243} $ & 11.9671\\
H5 & $9.259_{ -0.071}^{+0.141}  $ & $   -0.918_{-0.294}^{+0.222} $ & 5.1633\\
PAndAS-16 & $10.092_{ -0.102}^{+0.141}  $ & $   -0.918_{-0.772}^{+0.246} $ & 12.8005\\
PAndAS-17 & $9.884_{ -0.178}^{+0.143}  $ & $   -0.331_{-0.189}^{+0.302} $ & 4.9908\\
H18 & $10.3_{ -0.06}^{+0.0}  $ & $   -1.653_{-0.557}^{+0.393} $ & 2.0369\\
H25 & $10.3_{ -0.27}^{+0.0}  $ & $   -2.24_{-0.0}^{+0.77} $ & 1.38\\
H4 & $9.988_{ -0.222}^{+0.284}  $ & $   -1.506_{-0.442}^{+0.468} $ & 4.8489\\
PAndAS-06 & $9.78_{ -0.322}^{+0.177}  $ & $   -1.065_{-0.442}^{+0.285} $ & 1.1842\\
PAndAS-27 & $9.884_{ -0.373}^{+0.416}  $ & $   -1.065_{-0.385}^{+0.466} $ & 0.4242\\
PAndAS-44 & $9.467_{ -0.134}^{+0.301}  $ & $   -0.918_{-0.334}^{+0.277} $ & 2.032\\
G339 & $10.3_{ -0.667}^{+0.0}  $ & $   -1.065_{-1.051}^{+0.375} $ & 1.0768\\
G353 & $10.3_{ -0.058}^{+0.0}  $ & $   -1.359_{-0.72}^{+0.348} $ & 0.9364\\
H2 & $10.092_{ -0.51}^{+0.208}  $ & $   -2.093_{-0.147}^{+0.755} $ & 1.2174\\
PAndAS-37 & $9.988_{ -0.293}^{+0.312}  $ & $   -0.184_{-0.246}^{+0.266} $ & 1.5017\\
PAndAS-52 & $9.571_{ -0.304}^{+0.282}  $ & $   -0.625_{-0.509}^{+0.378} $ & 0.8456\\
H26 & $9.884_{ -0.627}^{+0.416}  $ & $   -1.653_{-0.587}^{+0.566} $ & 1.5356\\
  \enddata
\end{deluxetable}

\subsection{Fitting with {\sc ULySS}}

We employ the {\sc ULySS} \citep{2009A&A...501.1269K} package for fitting to derive the ages and metallicities of star clusters. The \cite{2010MNRAS.404.1639V} SSP models cover the wavelength ranges of 3540.5{\AA} --
7409.6{\AA} at a full width at half maximum (FWHM) of 2.3{\AA}.  The
models are based on the {\sc Miles} (Medium-resolution INT Library
of Empirical Spectra) spectral library \citep{2006MNRAS.371..703S}. The stellar
initial mass function (IMF) of \cite{1955ApJ...121..161S} is adopted for the
fitting and the solar-scaled theoretical isochrones of \cite{2000A&AS..141..371G}
have been used. The ranges of age and metallicity are $10^8$--$1.5\times10^{10}$ yr and [Fe/H] = $-2.32$ dex ($Z=0.0004$) -- $+0.22$ dex
($Z=0.03$) respectively. Furthermore, another independent SSP model, {\sc
  pegase-hr}, which is provided by \cite{2004A&A...425..881L}, is based on the
empirical spectral library {\sc Elodie} (e.g., \citealt{2001A&A...369.1048P}; \citealt{2007astro.ph..3658P}).
The wavelength coverage is 3900{\AA} -- 6800{\AA} with a spectral
resolution $R\sim10,000$. In this model, the fitted stellar parameters are effective temperature, $T_{\rm eff}$
(3100--50,000 K), gravity $log~g$ ($-0.25$ dex -- 4.9 dex), and
metallicity $\rm [Fe/H]$ ($-3$ dex -- $+1$ dex). The flux
calibration accuracy is 0.5--2.5\%. We adopt the {\sc pegase-hr} SSP
models with the \cite{1955ApJ...121..161S} IMF. The age ranges from $10^7$ to
$1.5\times10^{10}$ yr, and the metallicity [Fe/H] = $-2.0$ dex
($Z=0.0004$) to $+0.4$ dex ($Z=0.05$).

Table~5 presents the ages and metallicities derived from the full-spectrum fitting with {\sc ULySS} 
using both the \cite{2010MNRAS.404.1639V} and {\sc pegase-hr} models. Uncertainties were calculated using Monte-Carlo simulations to estimate biases, random errors, and parameter degeneracies. For each fit, random noise was injected into the data to generate a distribution of solutions, from which the final errors were determined.

\begin{deluxetable*}{r|cccc|cccc}
  \turnpage
  \tablecolumns{9} \tablewidth{0pc} \tablecaption{Ages, metallicities, and radial velocity derived from full-spectrum fitting with ULySS using the Vazdekis and {\sc pegase-hr} models.
    \label{t4.tab}}
  \tablehead{\colhead{} & \multicolumn{4}{c}{Vazdekis model} &\multicolumn{4}{c}{
      {\sc pegase-hr} model} \\
    \cline{2-9}
    \colhead{ID} & \colhead{log $t$} &  \colhead{$\rm
      [Fe/H]$} &  \colhead{Vr} & \colhead{$\chi^2_{min}/dof$}  & \colhead{log $t$} &
    \colhead{$\rm [Fe/H]$} &   \colhead{Vr} & \colhead{$\chi^2_{min}/dof$}  \\
    \colhead{} & \colhead{(yr)} & \colhead{(dex)} & \colhead{$km~s^{-1}$} & \colhead{} &
    \colhead{(yr)} & \colhead{(dex)} & \colhead{$km~s^{-1}$} & \colhead{}}
  \startdata
      EXT8  & $  9.96 \pm  0.06$  & $ -2.32 \pm 0.01$  & $    -93 \pm   16$  &  0.02 & $ 10.06 \pm  0.08$  & $ -2.30 \pm 0.01$  & $   -103 \pm   17$  &  0.02  \\
      MGC1  & $  9.81 \pm  0.12$  & $ -1.66 \pm 0.20$  & $   -211 \pm   22$  &  0.02 & $  9.79 \pm  0.08$  & $ -1.69 \pm 0.16$  & $   -201 \pm   24$  &  0.02  \\
 PAndAS-01  & $  9.94 \pm  0.58$  & $ -2.32 \pm 0.01$  & $   -186 \pm  108$  &  0.02 & $ 10.30 \pm  0.00$  & $ -2.30 \pm 0.01$  & $   -156 \pm  114$  &  0.02  \\
      B514  & $ 10.07 \pm  0.07$  & $ -1.92 \pm 0.08$  & $   -605 \pm   12$  &  0.01 & $ 10.27 \pm  0.01$  & $ -2.01 \pm 0.08$  & $   -605 \pm   13$  &  0.01  \\
      G002  & $  9.93 \pm  0.06$  & $ -2.14 \pm 0.07$  & $   -493 \pm   11$  &  0.01 & $ 10.02 \pm  0.09$  & $ -2.30 \pm 0.01$  & $   -502 \pm   12$  &  0.01  \\
        H1  & $ 10.07 \pm  0.11$  & $ -1.94 \pm 0.05$  & $   -204 \pm   12$  &  0.01 & $ 10.29 \pm  0.01$  & $ -1.98 \pm 0.07$  & $   -189 \pm   11$  &  0.01  \\
       H10  & $  9.90 \pm  0.06$  & $ -1.32 \pm 0.10$  & $   -340 \pm   12$  &  0.01 & $  9.86 \pm  0.05$  & $ -1.62 \pm 0.09$  & $   -344 \pm   12$  &  0.02  \\
 PAndAS-46  & $  9.91 \pm  0.06$  & $ -2.07 \pm 0.06$  & $    -69 \pm   10$  &  0.01 & $  9.89 \pm  0.04$  & $ -2.09 \pm 0.08$  & $    -64 \pm   10$  &  0.01  \\
 PAndAS-53  & $ 10.07 \pm  0.08$  & $ -1.92 \pm 0.04$  & $   -288 \pm    7$  &  0.01 & $ 10.02 \pm  0.05$  & $ -2.09 \pm 0.04$  & $   -276 \pm    6$  &  0.01  \\
 PAndAS-54  & $ 10.07 \pm  0.12$  & $ -1.94 \pm 0.06$  & $   -254 \pm   15$  &  0.01 & $ 10.02 \pm  0.09$  & $ -2.20 \pm 0.08$  & $   -249 \pm   14$  &  0.01  \\
      B517  & $  9.92 \pm  0.01$  & $  0.09 \pm 0.01$  & $    -92 \pm    1$  &  0.10 & $  9.85 \pm  0.01$  & $  0.11 \pm 0.01$  & $    -89 \pm    1$  &  0.11  \\
       H12  & $ 10.07 \pm  0.04$  & $ -1.97 \pm 0.08$  & $   -440 \pm   13$  &  0.01 & $  9.92 \pm  0.05$  & $ -2.12 \pm 0.08$  & $   -432 \pm   14$  &  0.01  \\
       H23  & $ 10.04 \pm  0.15$  & $ -1.22 \pm 0.16$  & $   -285 \pm   16$  &  0.02 & $  9.86 \pm  0.12$  & $ -1.18 \pm 0.15$  & $   -284 \pm   16$  &  0.02  \\
       H27  & $  9.93 \pm  0.12$  & $ -2.25 \pm 0.06$  & $   -386 \pm   12$  &  0.01 & $  9.90 \pm  0.04$  & $ -2.30 \pm 0.01$  & $   -387 \pm   13$  &  0.01  \\
        H5  & $ 10.12 \pm  0.20$  & $ -2.12 \pm 0.10$  & $   -499 \pm   31$  &  0.02 & $ 10.15 \pm  0.10$  & $ -2.22 \pm 0.16$  & $   -581 \pm   39$  &  0.02  \\
 PAndAS-16  & $  9.93 \pm  0.07$  & $ -2.32 \pm 0.01$  & $   -476 \pm   14$  &  0.02 & $ 10.03 \pm  0.10$  & $ -2.30 \pm 0.01$  & $   -474 \pm   14$  &  0.02  \\
 PAndAS-17  & $ 10.10 \pm  0.10$  & $ -0.77 \pm 0.07$  & $   -248 \pm   10$  &  0.02 & $  9.95 \pm  0.11$  & $ -0.64 \pm 0.10$  & $   -246 \pm   10$  &  0.02  \\
       H18  & $  9.95 \pm  0.16$  & $ -1.86 \pm 0.12$  & $   -369 \pm   20$  &  0.02 & $  9.88 \pm  0.10$  & $ -2.20 \pm 0.11$  & $   -366 \pm   22$  &  0.02  \\
       H25  & $ 10.06 \pm  0.09$  & $ -1.70 \pm 0.15$  & $   -280 \pm   15$  &  0.01 & $  9.96 \pm  0.10$  & $ -1.87 \pm 0.11$  & $   -274 \pm   13$  &  0.01  \\
        H4  & $  9.95 \pm  0.06$  & $ -1.80 \pm 0.14$  & $   -421 \pm   15$  &  0.01 & $  9.81 \pm  0.07$  & $ -1.82 \pm 0.15$  & $   -415 \pm   16$  &  0.01  \\
 PAndAS-06  & $  9.94 \pm  0.20$  & $ -2.20 \pm 0.09$  & $   -444 \pm   26$  &  0.01 & $ 10.03 \pm  0.12$  & $ -2.14 \pm 0.11$  & $   -436 \pm   26$  &  0.01  \\
 PAndAS-27  & $ 10.01 \pm  0.11$  & $ -1.31 \pm 0.13$  & $      5 \pm   14$  &  0.01 & $  9.78 \pm  0.09$  & $ -1.26 \pm 0.13$  & $      2 \pm   15$  &  0.01  \\
 PAndAS-44  & $  9.92 \pm  0.07$  & $ -2.32 \pm 0.01$  & $   -337 \pm   16$  &  0.01 & $  9.90 \pm  0.05$  & $ -2.30 \pm 0.01$  & $   -336 \pm   15$  &  0.01  \\
      G339  & $ 10.08 \pm  0.20$  & $ -1.80 \pm 0.12$  & $   -138 \pm   29$  &  0.01 & $ 10.17 \pm  0.03$  & $ -1.73 \pm 0.15$  & $   -156 \pm   27$  &  0.01  \\
      G353  & $  9.94 \pm  0.15$  & $ -2.12 \pm 0.12$  & $   -333 \pm   21$  &  0.01 & $  9.90 \pm  0.09$  & $ -1.93 \pm 0.17$  & $   -326 \pm   20$  &  0.01  \\
        H2  & $ 10.08 \pm  0.13$  & $ -1.92 \pm 0.10$  & $   -553 \pm   27$  &  0.01 & $ 10.30 \pm  0.00$  & $ -2.14 \pm 0.12$  & $   -541 \pm   28$  &  0.01  \\
 PAndAS-37  & $ 10.22 \pm  0.06$  & $ -0.81 \pm 0.06$  & $   -352 \pm   10$  &  0.02 & $ 10.30 \pm  0.00$  & $ -0.56 \pm 0.04$  & $   -347 \pm   10$  &  0.02  \\
 PAndAS-52  & $ 10.12 \pm  0.17$  & $ -1.55 \pm 0.17$  & $   -427 \pm   20$  &  0.02 & $ 10.19 \pm  0.03$  & $ -1.58 \pm 0.12$  & $   -433 \pm   18$  &  0.02  \\
       H26  & $  9.69 \pm  0.28$  & $ -0.08 \pm 0.42$  & $    363 \pm   92$  &  0.01 & $ 10.03 \pm  0.57$  & $ -0.54 \pm 0.47$  & $    293 \pm   86$  &  0.01  \\
  \enddata
\end{deluxetable*}

\begin{figure*}
  \centering
      \includegraphics[angle=0,scale=0.08888]{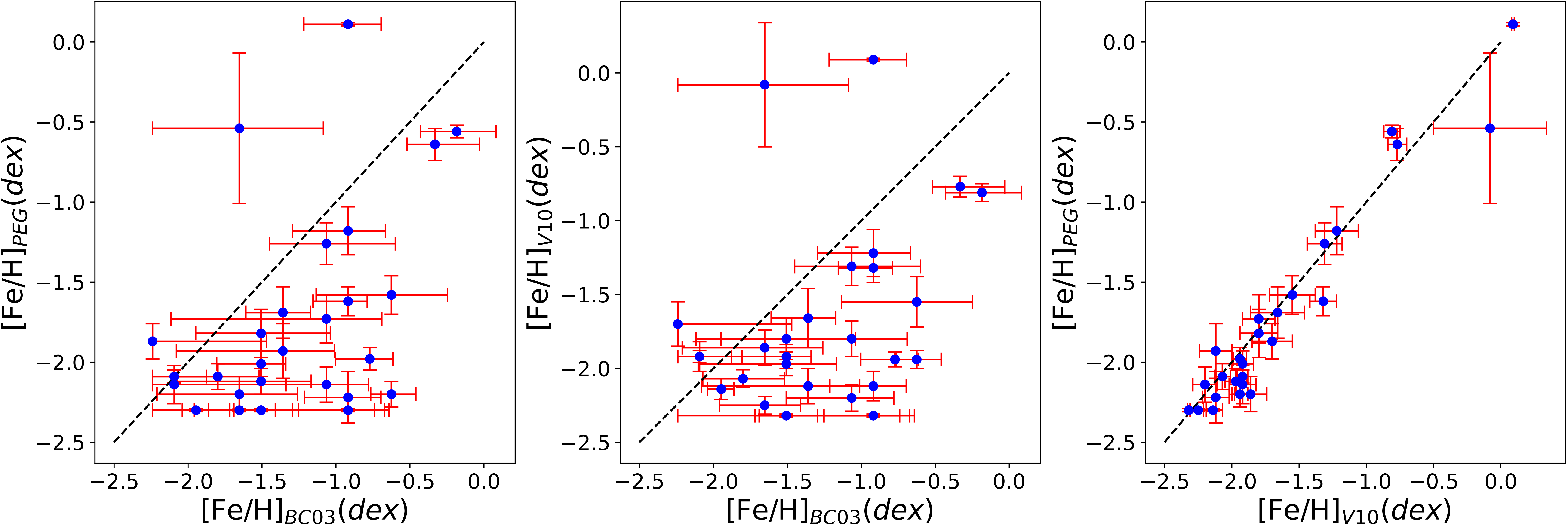}	
    \caption{Comparison of metallicities derived from different models. The x-axis shows results from the BC03 method, while the y-axis displays results from {\sc ULySS} using either the Vazdekis (V10) or {\sc pegase-hr} (PEG) models. The dashed line indicates a 1:1 relation. Error bars represent the $1\sigma$ uncertainties from each method.
    } 
  \label{fig1a}
\end{figure*}

\begin{figure*}
  \centering
      \includegraphics[angle=0,scale=0.08888]{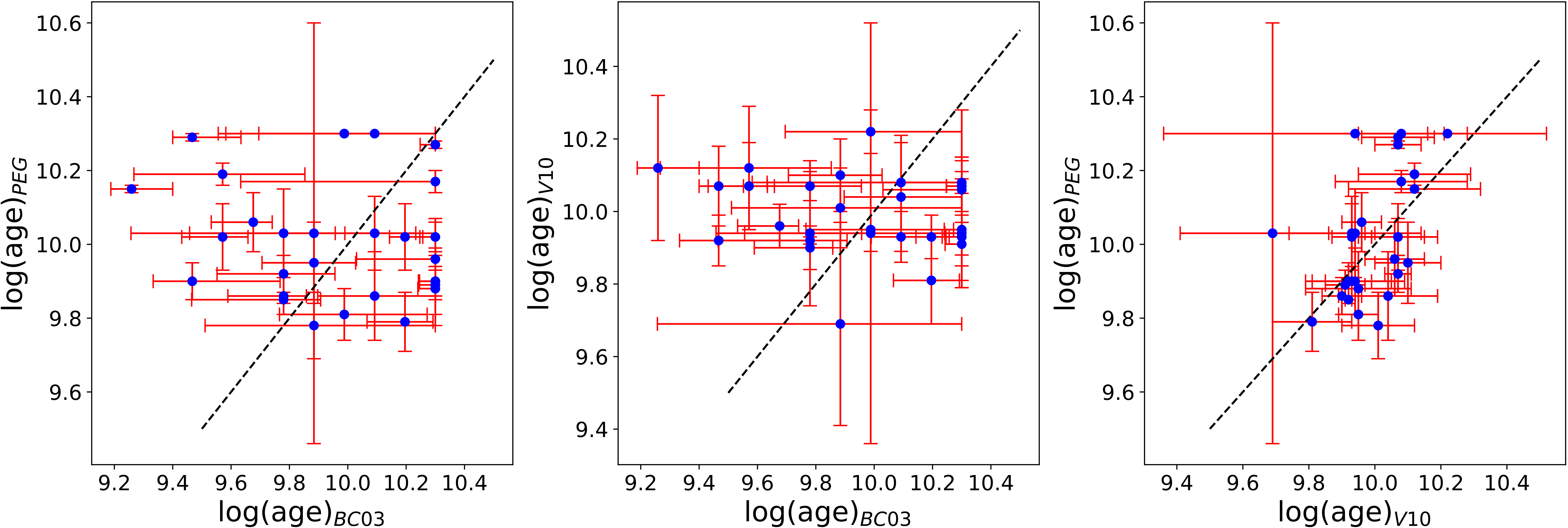}	
    \caption{Same as Figure~5, but for the derived ages.}
  \label{fig1a}
\end{figure*}

\subsection{Comparison between two method}
Figure~5 compares metallicities from the three methods. The two {\sc ULySS}-based measurements are consistent, whereas the metallicities from BC03 are generally higher compared to those from {\sc ULySS}. Given that the BC03 model spectra have relatively low spectral resolution and are not capable of accurately fitting individual metal absorption lines, and that the inclusion of photometric data further reduces the relative weight of spectral features in the fit, the model tends to choose templates that match the overall spectral energy distribution rather than reproducing detailed spectral features. Therefore, we considered the metallicity estimates from ULySS to be more reliable and adopted them as our preferred values. Except for cluster B517, where the poor spectral fit (indicated by a large $\chi ^2_\mathrm{red}$) necessitated using the BC03 value. 

Figure~6 compares the age estimates from the three methods. The  {\sc ULySS} results exhibit significant scatter, indicating a lower sensitivity to age. In contrast, the BC03 benefits from broader wavelength coverage, particularly the inclusion of UV and near-IR bands, which provides tighter constraints and smaller uncertainties for age. For this reason, we adopt the BC03 results as our final age estimates.

Table~6 summarizes the final adopted ages and metallicities, along with their uncertainties. Since the metallicity is derived from two ULySS models, we take the average of the two values as the final metallicity. The corresponding uncertainty is computed using equation~7, based on standard error propagation for the combination of two independent measurements.
The adopted age corresponds to the best-fit value from the BC03 model that minimizes the $\chi ^2$, and also constrained by the final metallicity derived from {\sc ULySS}.

\begin{deluxetable}{r|rr}
  \tablecolumns{3} \tablewidth{0pc} \tablecaption{Final ages and metallicities derived for the 29 M31 halo star clusters.}
  \tablehead{\colhead{ID} & \colhead{log $t$} &  \colhead{$\rm
      [Fe/H]$} \\
    \colhead{} & \colhead{(yr)} & \colhead{(dex)}} 
  \startdata
EXT8 & $10.018_{ -0.046}^{+0.04}  $ & $   -2.31\pm 0.007 $\\
MGC1 & $10.3_{ -0.017}^{+0.0}  $ & $   -1.675\pm 0.128 $\\
PAndAS-01 & $10.195_{ -0.203}^{+0.105}  $ & $   -2.31\pm 0.007 $\\
B514 & $10.3_{ -0.018}^{+0.0}  $ & $   -1.965\pm 0.057 $\\
G002 & $10.167_{ -0.025}^{+0.026}  $ & $   -2.22\pm 0.035 $\\
H1 & $10.3_{ -0.007}^{+0.0}  $ & $   -1.96\pm 0.043 $\\
H10 & $10.281_{ -0.052}^{+0.019}  $ & $   -1.47\pm 0.067 $\\
PAndAS-46 & $10.197_{ -0.031}^{+0.034}  $ & $   -2.08\pm 0.05 $\\
PAndAS-53 & $10.3_{ -0.002}^{+0.0}  $ & $   -2.005\pm 0.028 $\\
PAndAS-54 & $10.3_{ -0.002}^{+0.0}  $ & $   -2.07\pm 0.05 $\\
B517 & $9.87_{ -0.315}^{+0.127}  $ & $   -0.918_{ -0.299}^{+0.223} $\\
H12 & $10.047_{ -0.095}^{+0.101}  $ & $   -2.045\pm 0.057 $\\
H23 & $10.228_{ -0.14}^{+0.072}  $ & $   -1.2\pm 0.11 $\\
H27 & $10.165_{ -0.042}^{+0.047}  $ & $   -2.275\pm 0.03 $\\
H5 & $9.67_{ -0.056}^{+0.069}  $ & $   -2.17\pm 0.094 $\\
PAndAS-16 & $10.071_{ -0.087}^{+0.063}  $ & $   -2.31\pm 0.007 $\\
PAndAS-17 & $10.3_{ -0.037}^{+0.0}  $ & $   -0.705\pm 0.061 $\\
H18 & $10.261_{ -0.079}^{+0.039}  $ & $   -2.03\pm 0.081 $\\
H25 & $10.3_{ -0.037}^{+0.0}  $ & $   -1.785\pm 0.093 $\\
H4 & $10.03_{ -0.082}^{+0.123}  $ & $   -1.81\pm 0.103 $\\
PAndAS-06 & $10.3_{ -0.069}^{+0.0}  $ & $   -2.17\pm 0.071 $\\
PAndAS-27 & $10.252_{ -0.161}^{+0.048}  $ & $   -1.285\pm 0.092 $\\
PAndAS-44 & $10.3_{ -0.071}^{+0.0}  $ & $   -2.31\pm 0.007 $\\
G339 & $10.3_{ -0.025}^{+0.0}  $ & $   -1.765\pm 0.096 $\\
G353 & $10.257_{ -0.117}^{+0.043}  $ & $   -2.025\pm 0.104 $\\
H2 & $10.048_{ -0.258}^{+0.178}  $ & $   -2.03\pm 0.078 $\\
PAndAS-37 & $10.29_{ -0.07}^{+0.01}  $ & $   -0.685\pm 0.036 $\\
PAndAS-52 & $10.3_{ -0.028}^{+0.0}  $ & $   -1.565\pm 0.104 $\\
H26 & $9.035_{ -0.034}^{+0.03}  $ & $   -0.31\pm 0.315 $\\
  \enddata
\end{deluxetable}

\begin{equation}
\sigma _{[metal]}=\sqrt{\frac{\sigma ^2 _1+\sigma ^2 _2}{4}}
\end{equation}

\subsection{Comparison with other studies}
Given the large number of previous studies on the ages and metallicities of M31 globular clusters, it is essential to cross-match common sources, compare the performance of different methods, and assess the reliability of our results. For example, \cite{2016AJ....152...45C} (hereafter C16) derived ages and metallicities for M31 clusters by combining LAMOST spectra with SDSS photometric bands. However, their full spectral fitting approach was limited to a narrow wavelength range and did not incorporate ultraviolet or infrared data. Subsequently, \cite{2019A&A...623A..65W} (hereafter W19) used photometry with broader wavelength coverage—employing 15 medium- and narrow-band filters in the optical regime, which significantly increased the information content for age estimation while maintaining competitive metallicity precision. \cite{2021A&A...645A.115W} (hereafter W21) adopted a methodology similar to C16, but first applied machine learning to classify clusters by age, and then performed a more refined joint analysis combining LAMOST spectra and multi-band photometry according to the characteristics of each method. \cite{2024MNRAS.528.6010U} (hereafter C24) utilized high-resolution near-infrared spectra around the Ca\,II triplet lines to achieve highly precise metallicity measurements; however, their age estimates were only based on SDSS photometry, which has limited wavelength coverage and lower precision.

In this work, we build upon the broad wavelength coverage photometry similar to W19 and combine it with new low-resolution spectroscopic data obtained with the 2.16-meter telescope. This enables us to validate previous results and provide updated, more precise measurements of cluster parameters.

\begin{figure*}
  \centering
      \includegraphics[angle=0,scale=0.08888]{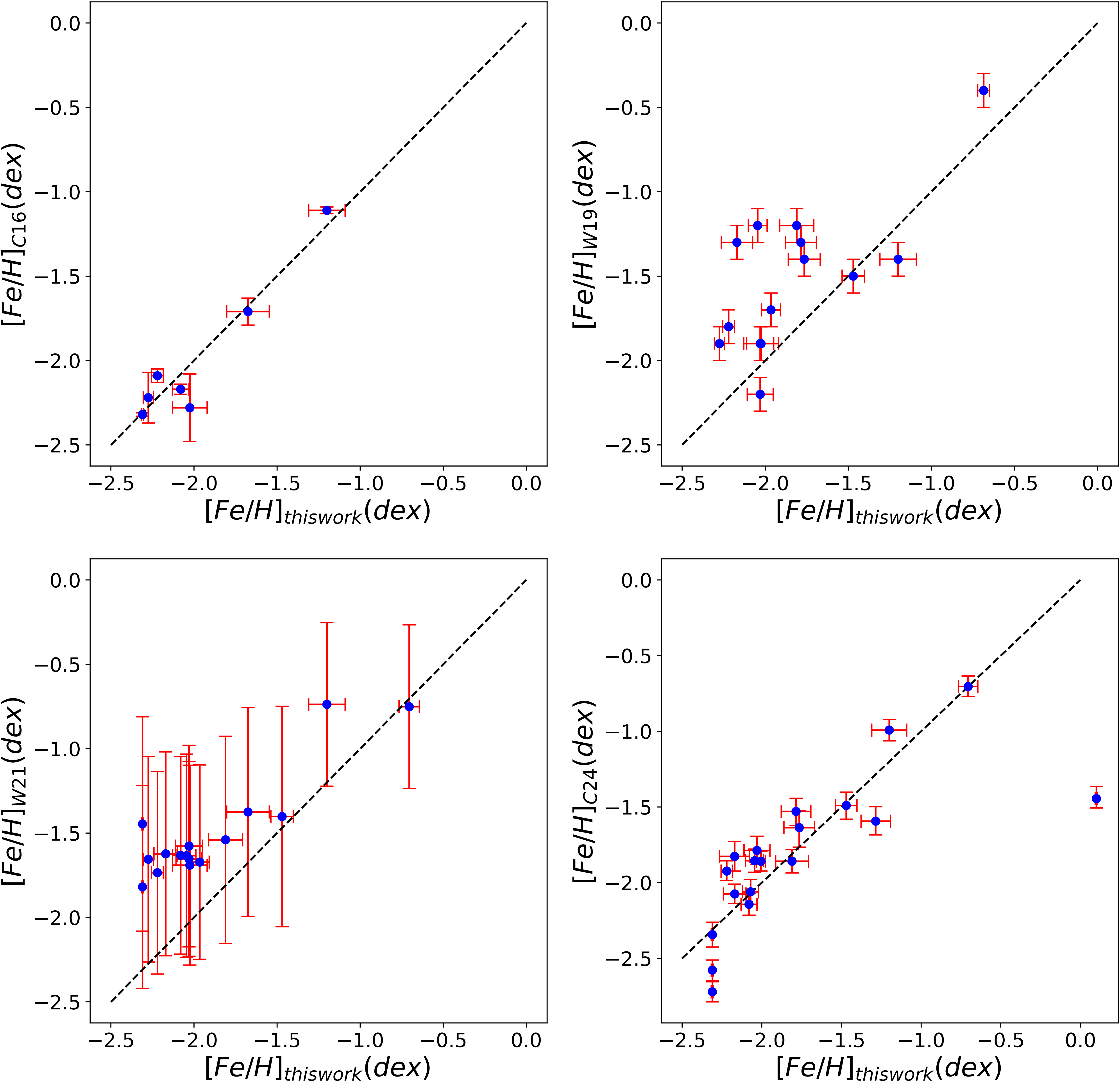}	
    \caption{Comparison of our final metallicity determinations (x-axis) with literature values (y-axis). The dashed line indicates the 1:1 relation. The literature sources are labeled as follows: C16: \citet{2016AJ....152...45C}; W19: \citet{2019A&A...623A..65W}; W21: \citet{2021A&A...645A.115W}; and C24: \citet{2024MNRAS.528.6010U}. Error bars represent the uncertainties reported in each study.
    } 
  \label{fig1a}
\end{figure*}
\begin{figure*}
  \centering
      \includegraphics[angle=0,scale=0.08888]{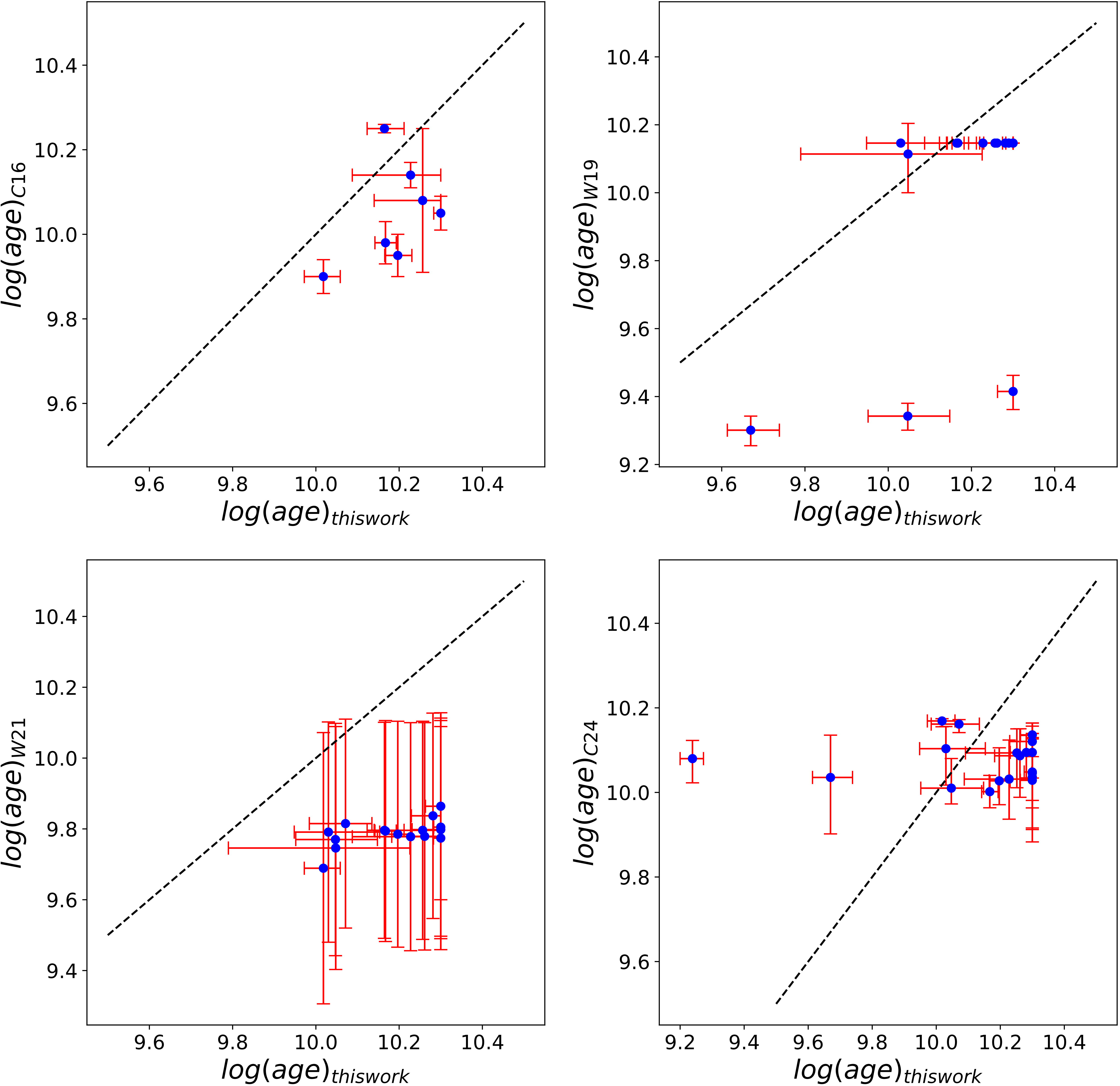}	
    \caption{Same as Figure~7, but  comparing our derived ages with literature values. }
  \label{fig1a}
\end{figure*}

Figure~7 compares the metallicities from this work with those from the four aforementioned studies. A good agreement is found between our results and those of C24, who used medium-resolution Ca II triplet spectra, as well as with C16, who relied on low-resolution LAMOST spectra. This consistency supports the robustness of our metallicity measurements. The larger scatter observed relative to W19 is likely due to differences in methodology; spectroscopic techniques typically offer greater sensitivity to metallicity than photometric methods alone.
For W21, a systematic offset is evident at [Z/H] $<$ −1.5. This discrepancy is also present in comparisons between W21 and both C16 and C24, indicating a potential systematic bias in the W21 analysis.

Figure~8 displays the age comparisons. A systematic difference is evident between our results and those of W21. Given that their inferred ages are predominantly below 10 Gyr, this discrepancy may arise from limitations in the stellar population models they used, such as boundary effects in the model grid. In contrast, our results show good overall agreement with the other three studies, despite some scatter and a few outliers. Closer agreement with W19 might be expected, as our photometry spans from the ultraviolet to the infrared, similar to their data.
However, we additionally include spectroscopic constraints on the SED shape, particularly in some key spectral regions.
These constraints appear to shift our results closer to the trends found in the ULySS-based study of C16. By combining the spectral resolution of spectroscopy with the broad wavelength coverage of photometry, our dataset maximizes the available information, theoretically yielding the most robust parameter estimates.

\section{Discussion}
\label{dis.sec}
Our sample consists of 29 star clusters in the M31 halo,
selected to be the brightest ($V<17.5$) and thus the most massive objects. Since the distance modulus
$(m-M)_0=24.64\pm0.15$, the absolute magnitude of our sample star
clusters is brighter than $M_0=-7.35$, corresponding to a mass exceeding $5.8\times10^4~\rm M_{\odot}$. 
Thus, our sample comprises massive star clusters and is not significantly biased by selection effects.
We have derived the ages and metallicities of our sample star
clusters using both {\sc ULySS} \citep{2009A&A...501.1269K} models, including the
\cite{2010MNRAS.404.1639V} and {\sc pegase-hr} SSP models, as well as the BC03 SSP models. While the BC03 lack ages $\log t<7$, and {\sc ULySS} models lack ages $\log t<8$ and $\log t<7$ yr for \cite{2010MNRAS.404.1639V} and {\sc pegase-hr} models, this does not impact our results as our youngest cluster is $\sim 10^9$ yr.

Since all clusters in our sample reside in the halo of M31, we neglect projection effects due to the disk inclination. Consequently, the projected galactocentric distance of each cluster depends only on its angular separation from the center of M31. We therefore adopt the J2000 coordinates of the M31 center as R.A. = 00h42m44.33s, Dec. = $+41^{\circ}16'07.5''$ from \href{https://simbad.u-strasbg.fr/simbad/}{SIMBAD Astronomical Database}, and a distance to M31 of 761 kpc from \cite{2021ApJ...920...84L}, to compute the physical projected distances of the 29 clusters from the M31 center. Our sample spans projected galactocentric distances from 25 to 120 kpc. The majority of these clusters are metal-poor ([Fe/H] $<$ −1.5) and old ($>$10 Gyr), consistent with the canonical properties of the M31 halo population. We do not observe the clear bimodal or multimodal metallicity distribution reported by \cite{2011AJ....141...61C}. This is likely due to our limited sample size and restriction to the outer halo, which may be insufficient to resolve a multi-Gaussian metallicity structure.
We do not observe a clear bimodal or multimodal metallicity distribution as anticipated by Caldwell et al. (2011). This may be due to the fact that our sample is confined entirely to the outer halo of M31 and is statistically limited, rendering it insufficient to resolve a multi-Gaussian metallicity structure. Nonetheless, a subset of globular clusters spatially associated with stellar substructures is discernible, exhibiting metallicities distinct from those of the majority of clusters unassociated with such features. This metallicity offset likely reflects an external origin—specifically, the accretion of dwarf satellite galaxies whose chemical enrichment histories differ from the in situ formation and evolutionary pathway of M31’s native halo population (e.g., \citealt{2019MNRAS.482.2795H}; \citealt{2022MNRAS.512.4819S}).

Regardless of whether substructure-associated clusters are included, we do not detect the negative metallicity gradient with increasing projected galactocentric distance that has been reported in prior studies (Figure~9). 
Quantitative assessment via linear regression—supplemented by both Pearson correlation coefficients and Spearman rank correlation tests—yields p-values far exceeding the conventional significance threshold of 0.05, indicating no statistically significant correlation between metallicity and projected radius. This might be because, although our sample spans a wide range in linear projected distance, its coverage in logR is relatively limited. Within these uncertainties, our findings remain consistent with those of \cite{2022MNRAS.512.4819S}. Nevertheless, our best-fit slope does not agree with the trend described in the eighth finding reported in the conclusion chapter of \cite{2014ApJ...780..128I}, which posits a clear decline in metallicity with radius in M31’s halo. From an alternative perspective, this result suggests that these halo globular clusters may originate from dwarf galaxies distinct from M31 and could even be the stripped nuclei of those systems, even though they do not appear to be associated with any known substructures in M31’s stellar halo. 

Among the 29 clusters, only three are younger than 10 Gyr (Figure~10). Two of them are spatially associated with known stellar substructures (Figure~11). B517 with [Fe/H] = −0.92 and age = 7.4 Gyr lies along the D stream, whose stellar population has metallicity ranging from -2.5 to −1.1 (\citealt{2019MNRAS.484.1756M}). Taking into account the error of metallicity, the cluster’s metallicity is broadly consistent with this stream. In contrast, H26, with [Fe/H] = −0.31 and age = 1.6 Gyr, overlaps with Stream C on the sky, but its metallicity is significantly richer than that of the stream itself, about [Fe/H] = −1.6 according to the work \cite{2014ApJ...780..128I}, suggesting that their alignment may be a projection effect rather than a physical association (\citealt{2014MNRAS.442.2929V}). The third young cluster, H5, is extremely metal-poor and shows no clear spatial correlation with any identified substructure.

Excluding H26, the most metal-rich clusters in our sample are PAndAS-17 and PAndAS-37, both with [Fe/H] about −0.7. PAndAS-17 is particularly noteworthy as the most metal-rich globular cluster in our dataset that is not associated with any identified substructure. This is consistent with \cite{2024MNRAS.528.6010U}, who describe PAndAS-17 as “the most metal-rich smooth halo globular cluster” and note that it is older than substructure-associated globular clusters of similar metallicity. PAndAS-37, on the other hand, is spatially aligned with the Giant Stellar Stream (GSS); its age and metallicity are also compatible with the GSS, indicating a likely physical connection.

\begin{figure*}
  \centering
      \includegraphics[angle=0,scale=0.12222]{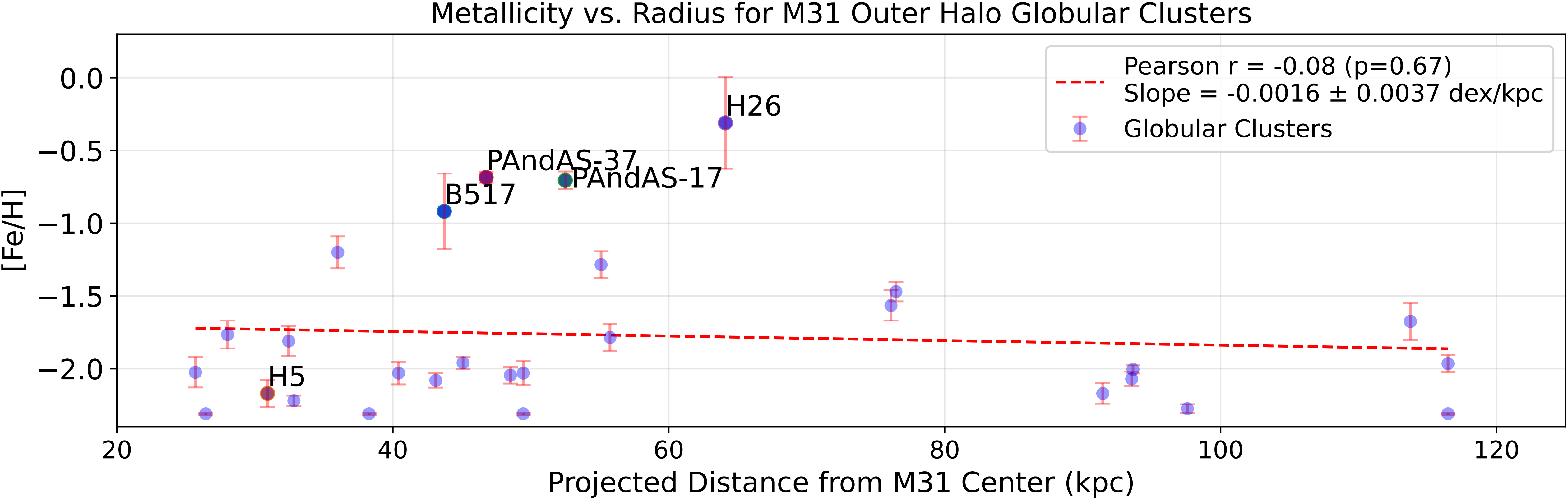}	
    \caption{Distribution of 29 globular clusters: projected distance versus metallicity. Red error bars represent the measurement uncertainties. The red dashed line indicates the linear fit, with the slope, Pearson coefficient (r), and p-value displayed in the legend.} 
  \label{fig1a}
\end{figure*}
\begin{figure*}
  \centering
      \includegraphics[angle=0,scale=0.12222]{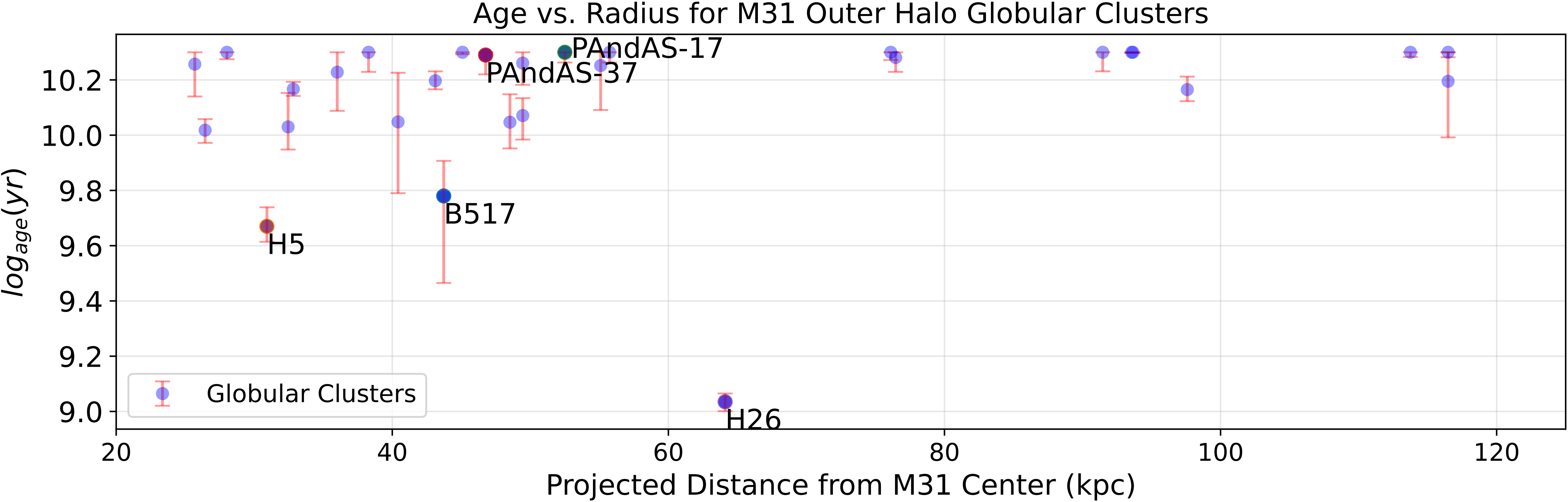}	
    \caption{Same as Figure~9, but for age versus projected galactocentric distance.} 
  \label{fig1a}
\end{figure*}
\begin{figure*}
  \centering
      \includegraphics[angle=0,scale=0.12222]{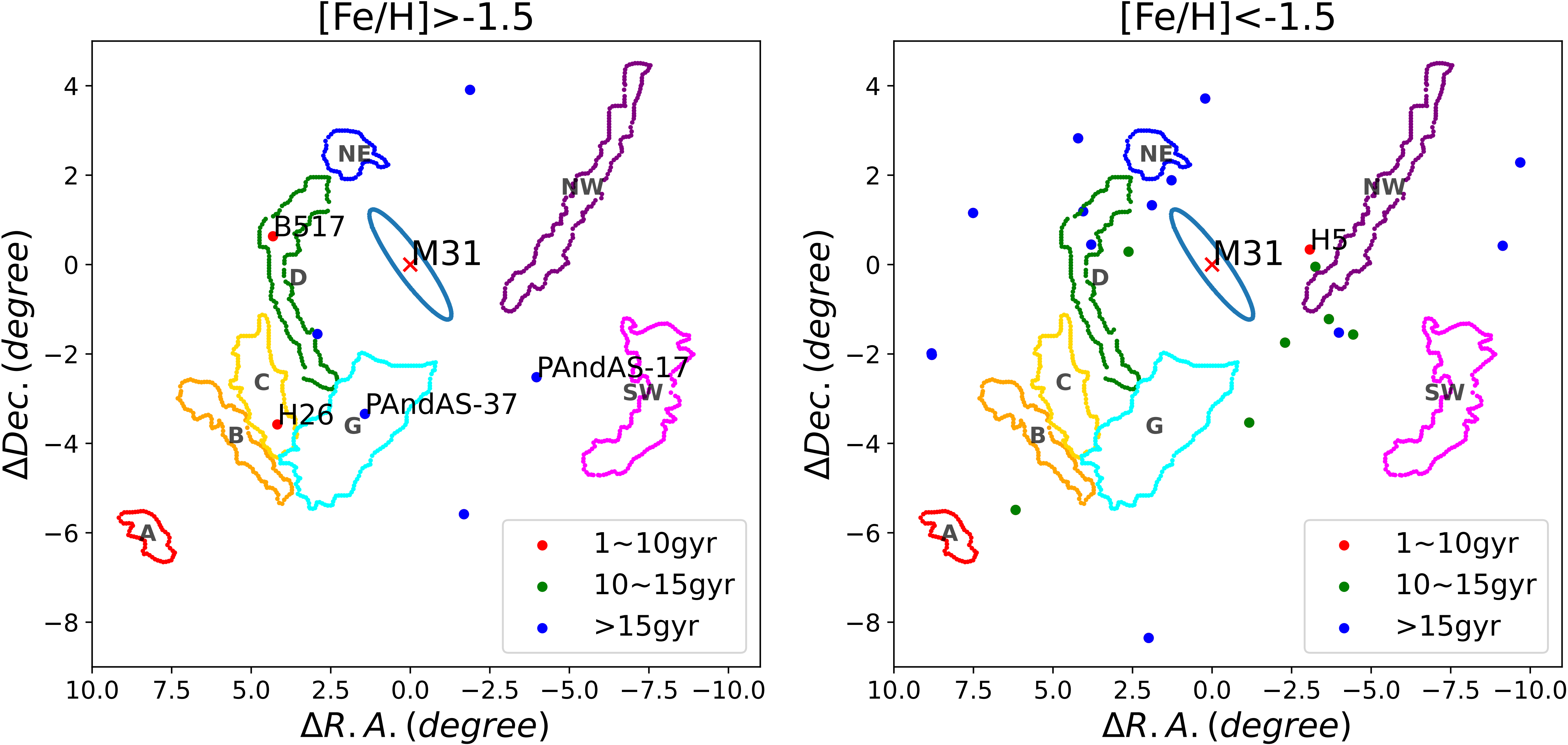}	
    \caption{Spatial distribution of 29 sample clusters overlaid on the M31 stellar substructure map from \citet{2010ApJ...717L..11M} (their Fig. 3). Substructures are labeled as follows: Streams A–-D; Giant Stream (G); NE structure (NE); NW Stream (NW); and SW Cloud (SW). The panels divide the sample by metallicity: metal-rich ([Fe/H] $>$ −1.5, left) and metal-poor ([Fe/H] ≤ −1.5, right). Symbol colors indicate age: red ($<$10 Gyr), green (10--15 Gyr), and blue ($>$15 Gyr). Labeled clusters are discussed in detail in the text.
    } 
  \label{fig1a}
\end{figure*}

\subsection{The UV-excess of our sample}

We compared the $GALEX$ FUV, NUV photometry with the SAGE $\rm
u_{S}$ band in relation to cluster ages. All the
magnitudes and colors were dereddened adopting the extinction
law of \cite{1989ApJ...345..245C} at $R_V=3.1$.

Excess emission in UV color has been observed in old populations,
e.g., discovered in early-type galaxies \citep{1976A&A....50..371D};
Galactic globular clusters: 47 Tuc \citep{1997AJ....114.1982O}, NGC 6388 and
NGC 6441 \citep{2007ApJS..173..643R}; and old open clusters: NGC 6791
\citep{2012ApJ...749...35B}. A similar phenomenon also has been modeled via
numerical N-body simulations \citep{2016RAA....16...37P}. 

In our relatively small sample, we did not detect any star clusters exhibiting UV excess at the current depth. To fully investigate this phenomenon among star clusters in the Local Group, the future
Chinese Space Station Telescope (CSST) will be an essential facility. The CSST will provide deep NUV photometry down to 255nm with a spatial resolution of $\sim0.15\arcsec$, approximately 10 times higher than that of $GALEX$ (\citealt{2018MNRAS.480.2178C}; \citealt{2019ApJ...883..203G}). This upcoming survey will enable a definitive assessment of the presence and properties of UV-excess clusters in M31.

\section{Summary and Conclusions}
\label{sum.sec}

In this work, we derived the physical parameters of 29 star clusters in the M31 halo using a combination of spectroscopic and photometric data. Our sample consists of bright clusters that lack previous detailed spectroscopic analysis. We obtained low-resolution spectra using the BFOSC spectrograph on the NAOC Xinglong 2.16-m telescope.

We compared three different fitting approaches: full-spectrum fitting using the {\sc ULySS} package \citep{2009A&A...501.1269K} including the
\cite{2010MNRAS.404.1639V} and  {\sc
  pegase-hr}  SSP models, and a joint spectroscopic-photometric analysis using the \cite{2003MNRAS.344.1000B} models (Padova1994 tracks with a Salpeter IMF \citep{1955ApJ...121..161S}). All fits were performed using $\chi^2$ minimization techniques.

Comparing our results with previous studies, we find general consistency for the clusters in common. We provide the first age and metallicity estimates for three of these clusters. For the majority of the sample, our work represents the first analysis to utilize joint fitting that spans from the ultraviolet to the infrared.

Although we do not detect a UV-excess in these M31 star clusters at current depths, we have demonstrated the robustness of combining spectroscopy with multi-band photometry to constrain stellar population parameters. We believe the cluster parameters provided here will serve as a valuable reference for future studies.

Finally, we highlight the synergy between large-scale surveys. The LAMOST survey \citep{2012RAA....12.1197C, 2012RAA....12..723Z} has provided the world's largest spectral dataset, with over ten million spectra. The SAGE Survey, with limiting magnitudes of $u_{SC}$ and $v_{SAGE}$ reaching $\sim17-18$ mag at SNR $\sim50$, perfectly matches the magnitude range of high-SNR LAMOST targets. Consequently, SAGES photometry offers a powerful complement to LAMOST spectroscopy for future stellar population studies.

\begin{acknowledgments}
We acknowledgment final support from the National Natural Science Foundation of China (NSFC) under grant No.12588202, National Key R\&D Program of China No.2023YFE0107800, No.2024YFA1611900, Strategic Priority Research Program of Chinese Academy of Sciences, grant No.1160102.; Sino-German Center Project GZ 1284; National Key Research and Development Program of China grant Nos. 2019YFA0405502, 2016YFA0400804. X.Y.P. expresses gratitude for support from the Research Development Fund of Xi'an Jiaotong Liverpool University (RDF-18-02-32) and the financial support of two grants of National Natural Science Foundation of China, Nos. 11673032 and 11503015. Juanjuan Ren thanks the support of the China Manned Space Program (Grant No. CMS-CSST-2025-A19). This work also supported by the Strategic Priority Research Program of the Chinese Academy of Sciences , Grant No. XDB0550100. This work is sponsored by the Xinjiang Uygur Autonomous Region "Tianchi Talent" Introduction Plan.
\end{acknowledgments}


\end{CJK*}
\end{document}